\documentclass[12pt,preprint]{aastex}
\usepackage{graphicx}

\newcommand{\La}{\mbox{${\rm Ly\alpha}$}}

\newcommand{\Hb}{\mbox{${\mathrm{H\beta}}$}}
\newcommand{\Hg}{\mbox{${\mathrm{H\gamma}}$}}

\newcommand{\Line}[3]{\Ion{#1}{#2}\,$\lambda$\,#3}
\newcommand{\Lines}[3]{\Ion{#1}{#2}\,$\lambda\lambda$\,#3}
\newcommand{\Ion}[2]{#1{\,\scriptsize #2}}
\newcommand{\Mwd}{\mbox{$M_{\rm wd}$}}
\newcommand{\Rwd}{\mbox{$R_{\rm wd}$}}
\newcommand{\Teff}{\mbox{$T_{\rm eff}$}}
\newcommand{\Porb}{\mbox{$P_{\rm orb}$}}

\newcommand{\Msun}{\mbox{$M_{\odot}$}}
\newcommand{\kms}{\mbox{$\rm km\,s^{-1}$}}
\shortauthors{G\"ansicke et al.}
\shorttitle{The accreting white dwarfs in BW\,Scl, BC\,UMa and SW\,UMa}

\begin{document}

\title{Hubble Space Telescope STIS Observations of the Accreting White
  Dwarfs in BW\,Scl, BC\,UMa and SW\,UMa\altaffilmark{1}}

\author{Boris T. G\"ansicke}
\affil{Department of Physics, University of Warwick, Coventry CV4 9BU}
\email{Boris.Gaensicke@warwick.ac.uk}

\author{Paula Szkody}
\affil{Astronomy Department, University of Washington, Seattle, WA 98195}
\email{szkody@astro.washington.edu}

\author{Steve B. Howell}
\affil{WIYN Observatory \& NOAO, 950 N. Cherry Avenue, Tucson, AZ  85726}
\email{howell@noao.edu}

\author{Edward M. Sion}
\affil{Department of Astronomy and Astrophysics, Villanova University, Villanova, PA 19085}
\email{Edward.Sion@villanova.edu}


\altaffiltext{1}{Based on
  observations made with the NASA/ESA Hubble Space Telescope obtained
  at the Space Telescope Science Institute which is operated by the
  AURA under NASA contract NAS 5-26555 and with the Apache Point
  Observatory (APO) 3.5-m telescope which is operated by the
  Astrophysical Research Consortium (ARC).}

\begin{abstract}
We have observed the short-period dwarf novae BW\,Scl, BC\,UMa and
SW\,UMa using the \textit{Hubble Space Telescope}/Space Telescope
Imaging Spectrograph. In all three systems, the white dwarf is the
dominant source of far-ultraviolet flux, even though in BC\,UMa and
SW\,UMa an additional continuum component contributes $\sim10$\,\% and
$\sim20$\,\% of the 1400\,\AA\ flux, respectively. Fitting the data
with detailed white dwarf model spectra, we determine the effective
temperatures to be $14\,800\pm900$\,K (BW\,Scl), $15\,200\pm1000$\,K
(BC\,UMa), and $13\,900\pm900$\,K (SW\,UMa). The additional continuum
component in BC\,UMa and SW\,UMa is equally well described by either a
blackbody or a power law, which could be associated with emission from
the hot spot or from an optically thin accretion disk (or an optically
thin layer on top of a colder optically thick disk),
respectively. Modelling the narrow metal lines detected in the STIS
spectra results in sub-solar abundances of carbon, oxygen and silicon
for all three systems, and also suggests substantial supra-solar
abundances of aluminium. The narrow absorption line profiles imply low
white dwarf rotation rates, $v\sin i\la300$\,\kms\ for the three white
dwarfs.  SW\,UMa is the only system that shows significant short-term
variability in the far-ultraviolet range, which is primarily
associated with the observed emission lines.
\end{abstract}

\section{Introduction}
The study of accreting white dwarfs in close binaries is best
accomplished using the far-ultraviolet (FUV) portion of the spectrum
for systems with low mass transfer rates. In these cases, the peak
flux of the 12\,000--50\,000K white dwarf \citep{sion99-1,
gaensicke00-1} dominates the accretion disk. If the spectral
resolution is on the order of 1\,\AA\ or better and the
signal-to-noise ratio (S/N) at least of the order of 10, then the
temperature, rotation rate and chemical abundances of the white dwarf
can be determined reliably from a single FUV spectrum.  This has been
demonstrated within our past observations and analyses using the Space
Telescope Imaging Spectrograph (STIS) for EK\,TrA
\citep{gaensickeetal01-3}, EG\,Cnc, HV\,Vir \citep{szkodyetal02-3},
LL\,And, EF\,Peg, \citep{howelletal02-1}, GW\,Lib
\citep{szkodyetal02-4}, VY\,Aqr, WX\,Cet \citep{sionetal03-1}, and
AL\,Com \citep{szkodyetal03-1}. The results of these past observations
have shown that the coolest white dwarfs in disk-accreting systems
cluster about 12\,000--14\,000K, compared to single field white
dwarfs, which have temperatures down to 5000\,K. Thus, the white
dwarfs in CVs are definitely heated by accretion \citep{sion95-1,
townsley+bildsten03-1}. However, it is clear that we do not yet fully
understand the relation between close binary evolution, accretion
heating and white dwarf cooling. The measured white dwarf temperatures
do not monotonically decrease with decreasing orbital period, which
would be expected in a simple picture where CVs evolve from long to
short orbital periods and from high to low mass transfer rates
\citep{howelletal01-1}. At least one additional parameter that
determines the white dwarf temperatures in CVs are the white dwarf
masses~--~which are poorly known.  Furthermore, although several CVs
are cool enough to lie in the instability strip for pulsating ZZ\,Ceti
white dwarfs (e.g. EG\,Cnc, \citealt{szkodyetal02-3}), only one of our
STIS targets (GW\,Lib) is known to pulsate~--~and paradoxically, its
white dwarf is apparently too hot to be a pulsator
\citep{szkodyetal02-4}. Addressing the complex interplay between white
dwarf properties and CV evolution clearly needs high-quality
observational input for a sufficiently large sample of CV white
dwarfs. In this paper, we present the analysis of the last three
short-period dwarf novae from our Cycle\,8 \textit{HST}/STIS program
displaying a white dwarf dominated far-ultraviolet spectrum: BW\,Scl,
BC\,UMa and SW\,UMa.

BW\,Scl was initially found in the Hamburg/ESO quasar survey as
HE\,2350-3908 \citep{augusteijn+wisotzki97-1} and independently
discovered in the ROSAT survey as RX\,J2353.0-3852
\citep{abbottetal97-1}.  Its extremely short orbital period
(78.2\,min), double-humped optical light curve (peak-to-peak variation
0.1\,mag), spectrum with double-peaked Balmer emission flanked by
broad absorption lines, and lack of visibility of any secondary star
led Augusteijn \& Wisotzki to conclude that this system was similar to
WZ,\,Sge a prototype of short period, low accretion-rate cataclysmic
variables with extreme amplitude outbursts that only occur on
time scales of tens of years. They used the X-ray flux to obtain a quiescent
accretion rate of only 10$^{-12}$M$_{\odot}$ yr$^{-1}$. To date, no
outburst of BW\,Scl has been observed.

BC\,UMa was first identified as a large amplitude dwarf nova
\citep{romano63-1}.  The first spectra by \citet{mukaietal90-1}
revealed evidence of the white dwarf from absorption surrounding the
emission lines, as well as a secondary star later than M5. They
accomplished a crude fit to the blue spectrum and obtained a
temperature of near 15\,000\,K and used the secondary to estimate a
distance of 130--400 pc.  Photometry by \citet{howelletal90-1}
revealed $B$ and $R$ light curves with a single peak-to-peak variation
of 0.25\,mag at a period of 91\,min when the system was at $V=18.6$. Later
photometry during quiescence and superoutburst \citep{pattersonetal03-1}
showed a double-humped light curve with smaller amplitude when the
system was at $V=17.4$. They also refined the period from spectroscopy
to 90.2\,min.

SW\,UMa is also a large amplitude, short period dwarf nova, but which
is known to have several types of outbursts
\citep{howelletal95-1}. The first time-resolved study
\citep{shafteretal86-2} revealed an orbital period of 81.8\,min from
the spectra, while the photometry showed a repeatable 0.4\,mag hump on
most nights. But on one night, when the mean magnitude was about
0.5\,mag fainter, the orbital variation disappeared and was replaced
by a quasi-periodic oscillation with a period near 16 min.  This
change from hump presence to absence was also evident in the datasets
of \citet{howell+szkody88-1}. Another peculiarity of this system is
the presence of a narrow H$\alpha$ emission component which is most
prominent at orbital phase 0.3. Doppler tomograms constructed from
several of the emission lines showed a prominent disk contribution in
the optical with enhanced emission at phases 0.3, 0.6 and 0.9
\citep{szkodyetal00-3}. Thus, SW\,UMa appears to have a higher mass
transfer rate and more extensive disk than BW\,Scl and
BC\,UMa. Despite this, \citet{gaensicke+koester99-1} were able to see
evidence of the white dwarf in the IUE spectra, and determined a
temperature for the white dwarf of 16\,000$\pm$1500K.

Our STIS spectra of these 3 systems are used to provide good constraints on
the white dwarf temperature, disk contributions and variability of the UV
compared to the optical light. Additionally, the
good S/N for the brighter systems BC\,UMa and SW\,UMa allows us to obtain
good values for the rotation rates and compositions of the white dwarfs. 

\section{Observations}
\subsection{BW\,Scl} 
BW\,Scl was observed in deep quiescence with \textit{HST} during a
single spacecraft orbit on 12 September 1999
(Table\,\ref{t-obslog}). From the STIS CCD acquisition image, we derive a
F28$\times$50LP magnitude of 16.5, roughly equivalent to an $R$ band
measurement.  The data were obtained using the E140M echelle grating
and the $0.2\arcsec\times0.2\arcsec$ aperture. The nominal spectral
resolution of this setup is $R\sim90000$, covering the range
1125$-$1710\,\AA. BW\,Scl was among the first targets observed in our
Cycle\,8 STIS program, and we opted for the echelle grating in order
to accurately measure the rotational velocity of the white dwarf from
the Doppler broadening of narrow metal lines. Unfortunately, the FUV
flux of BW\,Scl was lower than expected, and the quality of the data
is insufficient to exploit the high resolution of the E140M grating.
Figure\,\ref{f-stis_opt_spectra} shows the STIS data of BW\,Scl
rebinned to a resolution of $\sim250$\,\kms.  Clearly visible is
the broad flux turn over at $\lambda\la1400$\,\AA\, which we identify
as the \La\ absorption line from the white dwarf photosphere. In
addition, a large number of narrow metal absorption lines are
detected, the most prominent ones being \Lines{Si}{II}{1260,65}, a
complex of \Ion{C}{I} lines near 1280\,\AA, the blend of
\Lines{Si}{II}{1304,05,09} and \Lines{O}{I}{1304,06},
\Line{C}{II}{1335}, a feature at 1432\,\AA\ which is most likely due to
\Ion{Ca}{II} (possibly blended with some \Ion{C}{I} lines),
\Lines{Si}{II}{1527,33}, a blend of \Ion{C}{I} lines at 1556--58\,\AA,
and \Line{Al}{II}{1670}. Broad \La\ and \Line{C}{IV}{1550} emission are
clearly present, the detection of narrow \Line{C}{III}{1176} and
\Line{He}{II}{1640} emission is less secure. The quality of the data
decreases towards the blue and red ends of the spectrum because of the
decreasing response of the instrument. At the red end, the quality of
the spectrum is deteriorated due to problems in extracting the
individual echelle orders of this low-flux object, resulting in
spurious emission/absorption edges. Finally, three small gaps between
the echelle orders are present at 1653\,\AA, 1671\,\AA, and 1690\,\AA.

In order to extend the wavelength coverage of BW\,Scl for our analysis
we have extracted the optical spectrum  presented by
\citet{abbottetal97-1}, renormalizing it to the average magnitude of
the system, $V\simeq16.4$ (Fig.\,\ref{f-stis_opt_spectra}). The broad
Balmer absorption lines characteristic of a high-gravity white dwarf
photosphere are clearly discerned.

\subsection{BC\,UMa\label{s-obs_bcuma}}
\begin{figure}
\begin{minipage}[t]{8.7cm}
\includegraphics[angle=270,width=8.5cm]{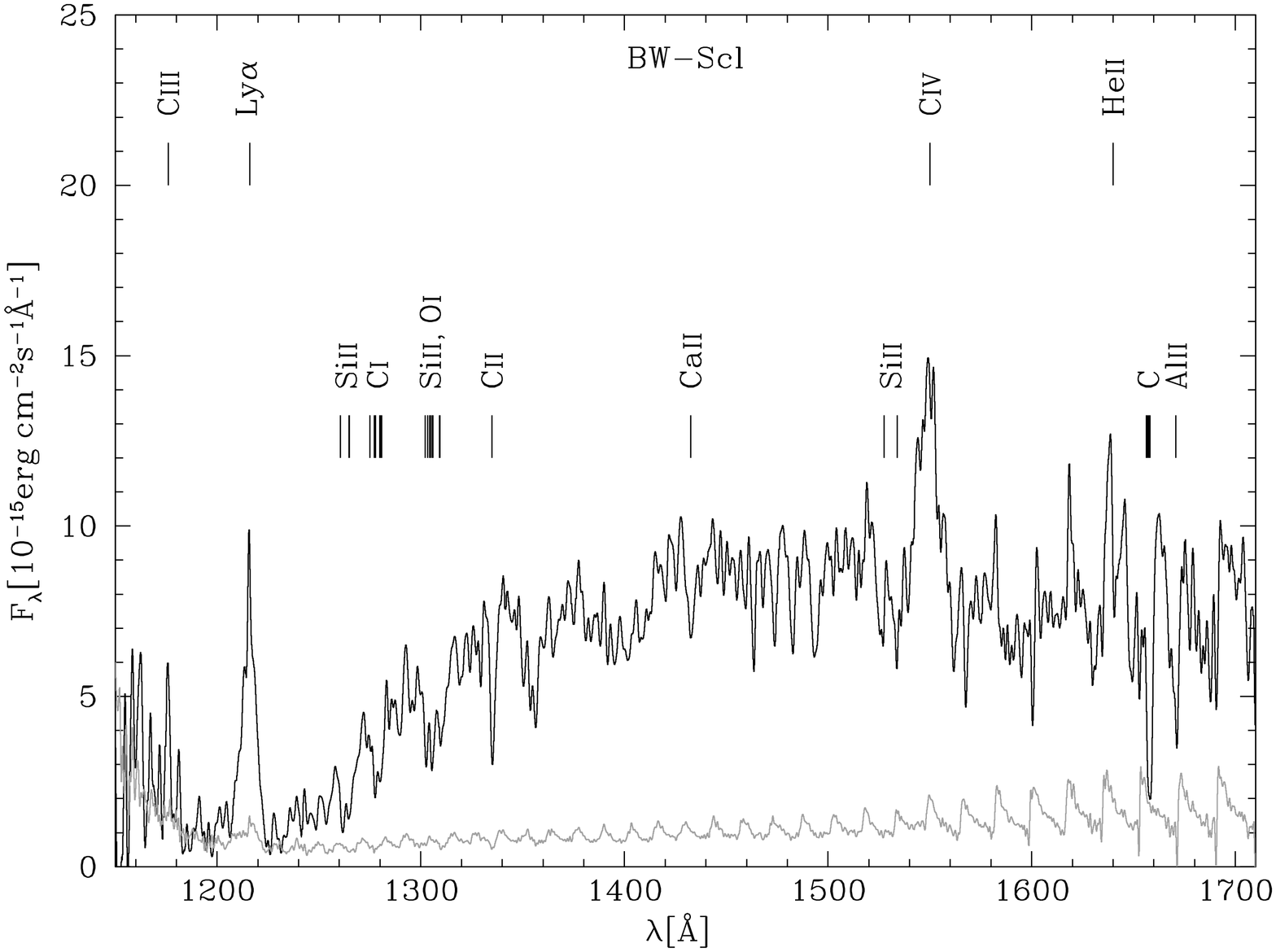}
\includegraphics[angle=270,width=8.5cm]{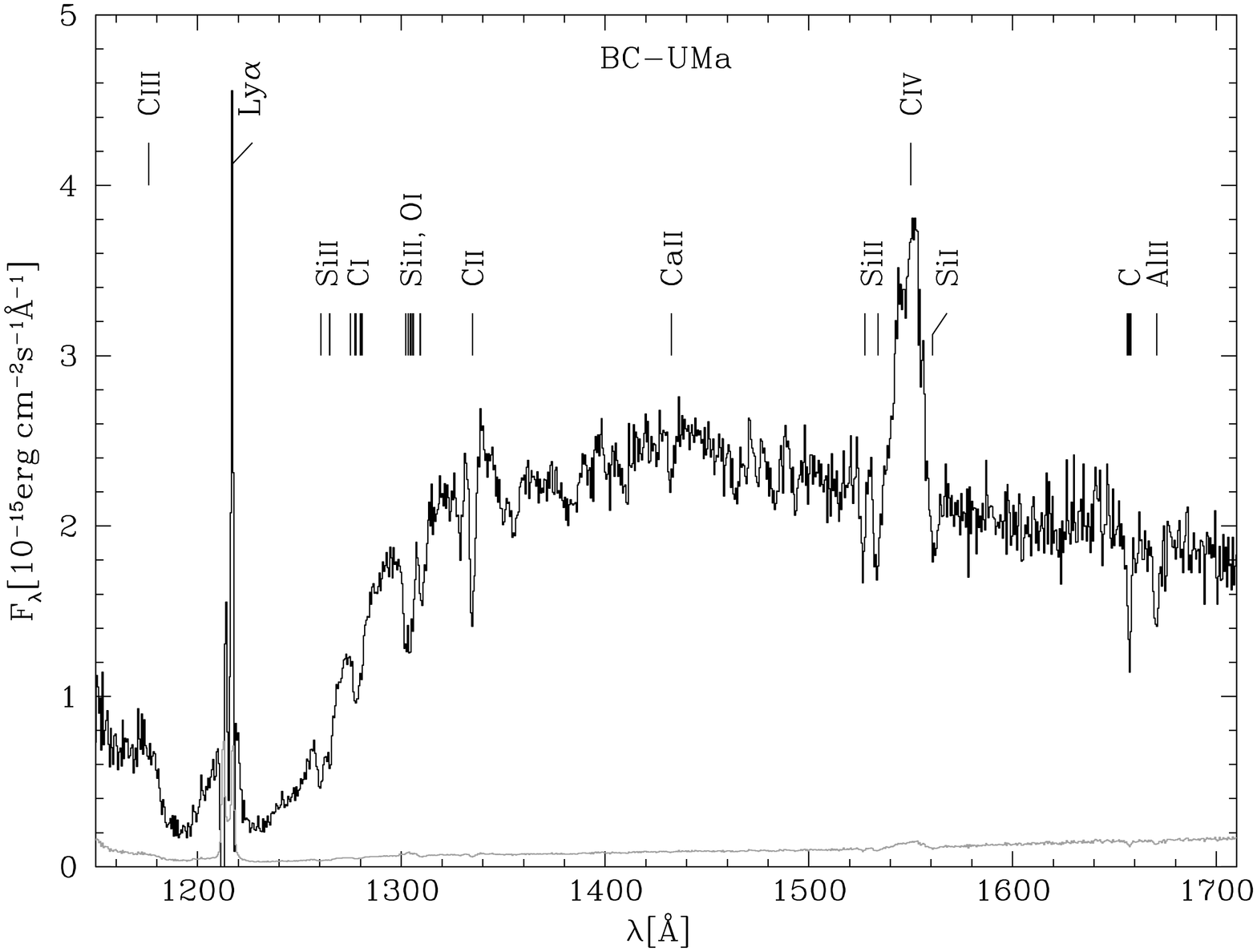}
\includegraphics[angle=270,width=8.5cm]{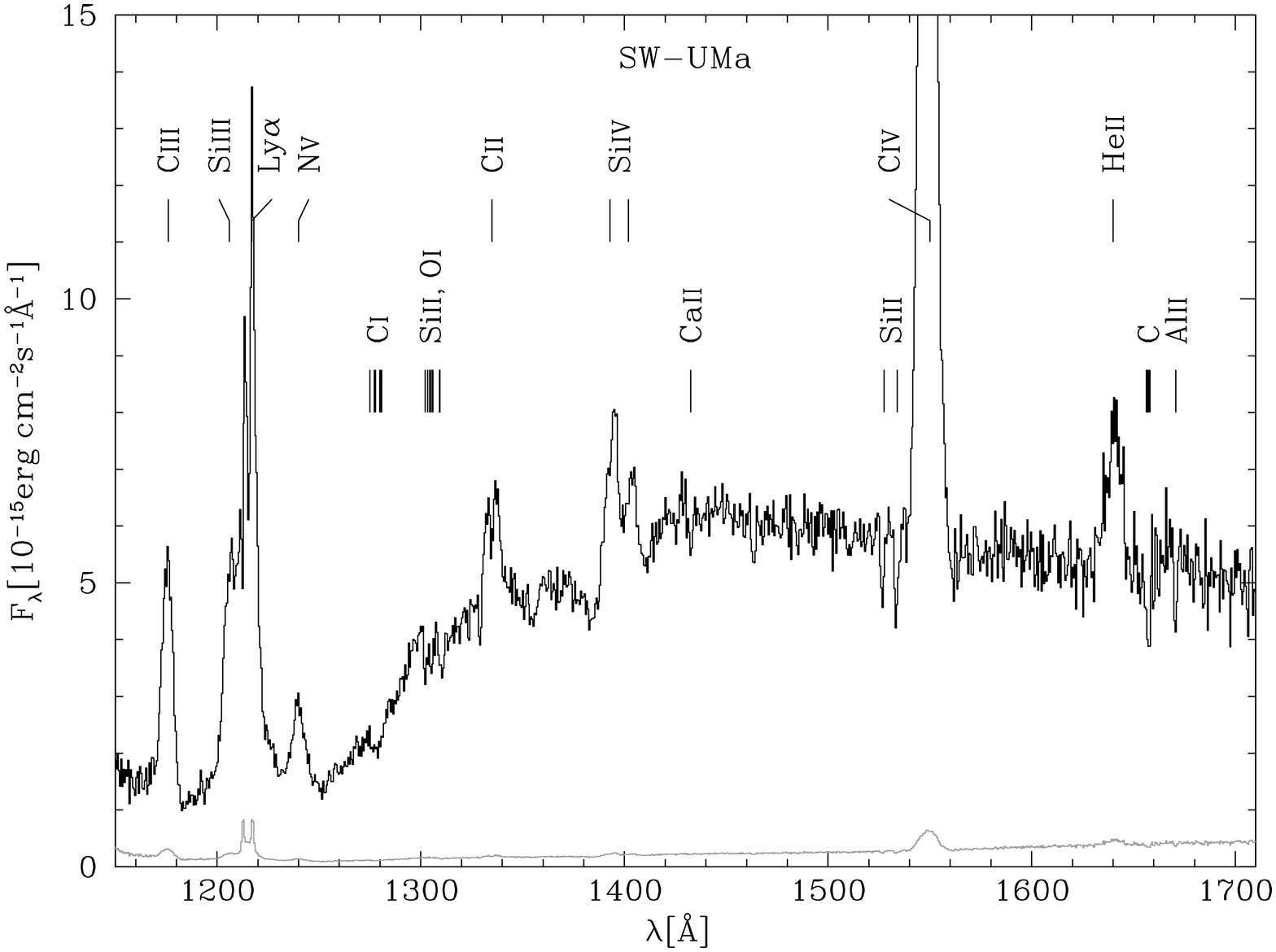}
\end{minipage}
\hfill
\begin{minipage}[t]{8.7cm}
\includegraphics[angle=270,width=8.5cm]{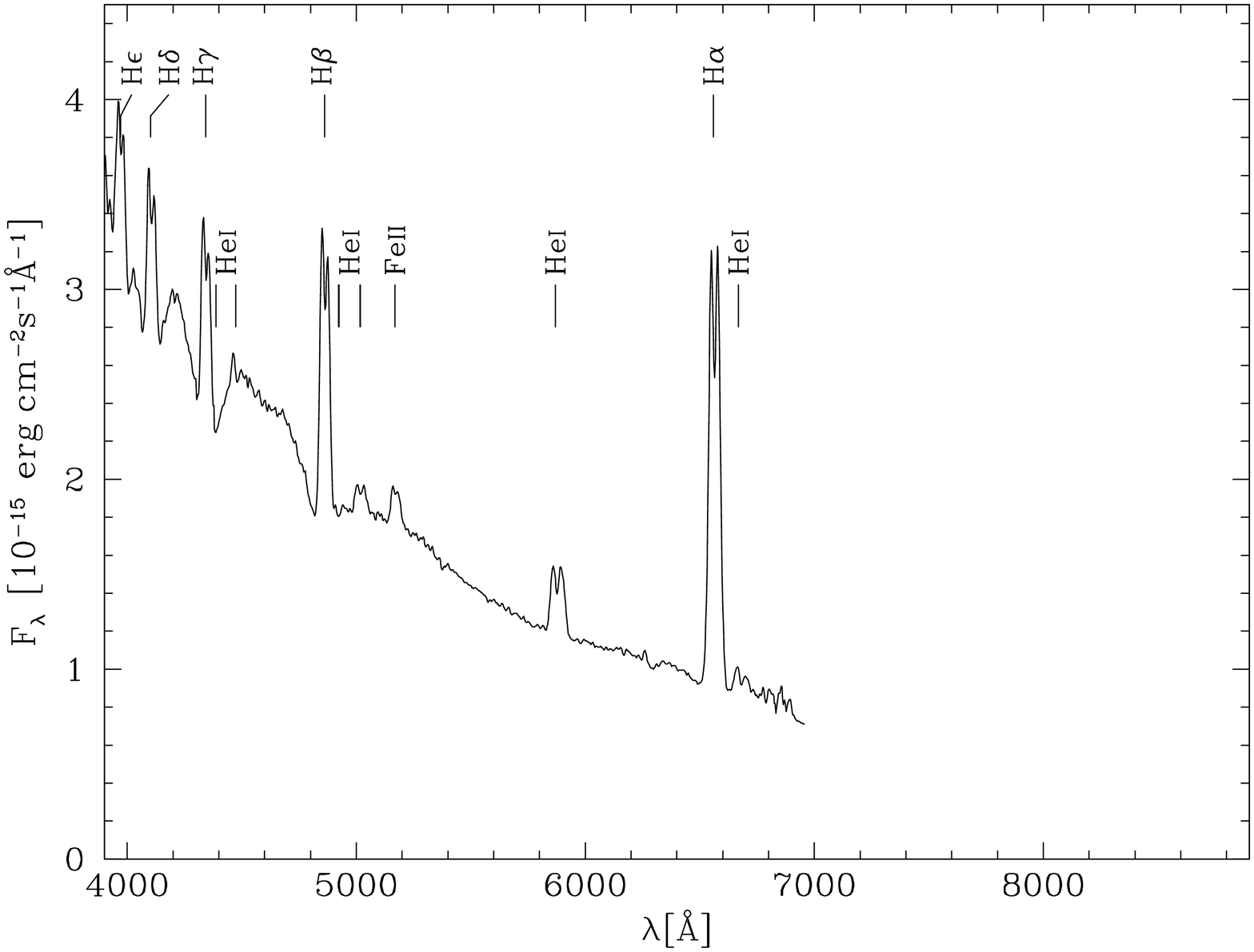}
\includegraphics[angle=270,width=8.5cm]{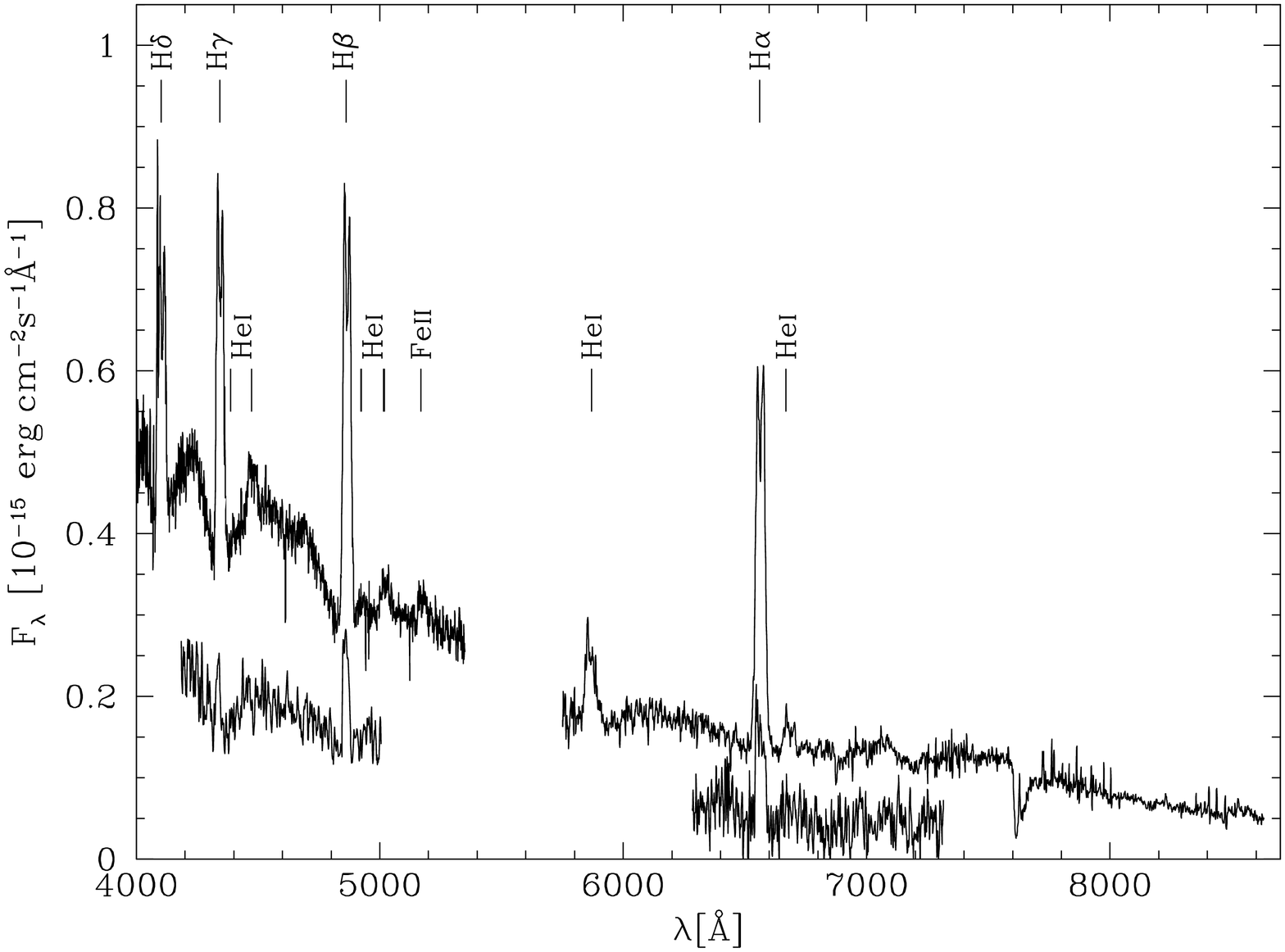}
\includegraphics[angle=270,width=8.5cm]{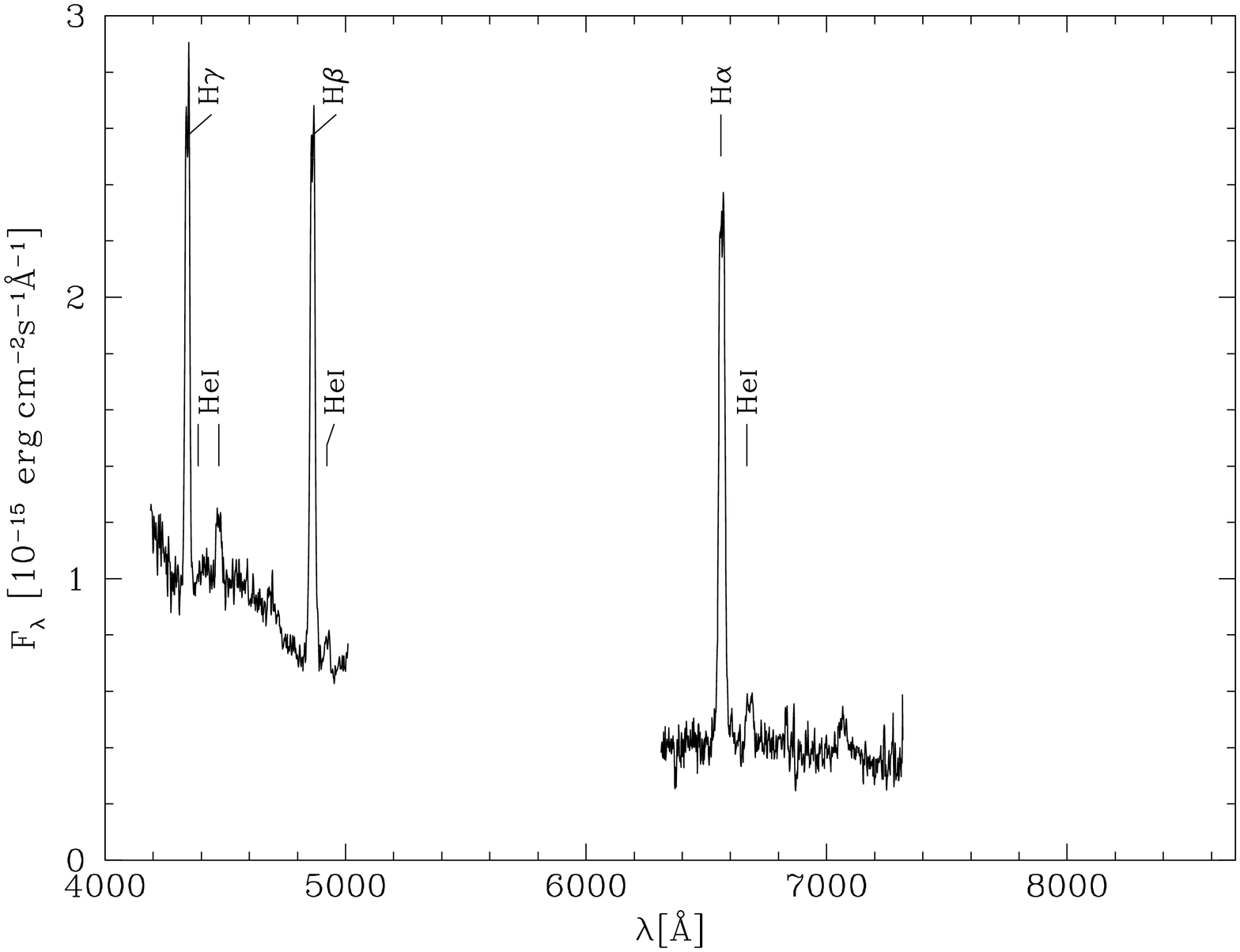}
\end{minipage}
\caption[]{\label{f-stis_opt_spectra}
Average \textit{HST}/STIS spectra of BW\,Scl (top), BC\,UMa (middle)
and SW\,UMa (bottom). Prominent emission/absorption lines are
identified. $1\sigma$ errors are shown in gray. BW\,Scl was observed
with the E140M grating, the spectrum shown here has been convolved
with a 0.6\,\AA\ Gaussian. The S/N varies slightly
over each individual echelle order.}
\end{figure}

\begin{figure}
\includegraphics[angle=270,width=12cm]{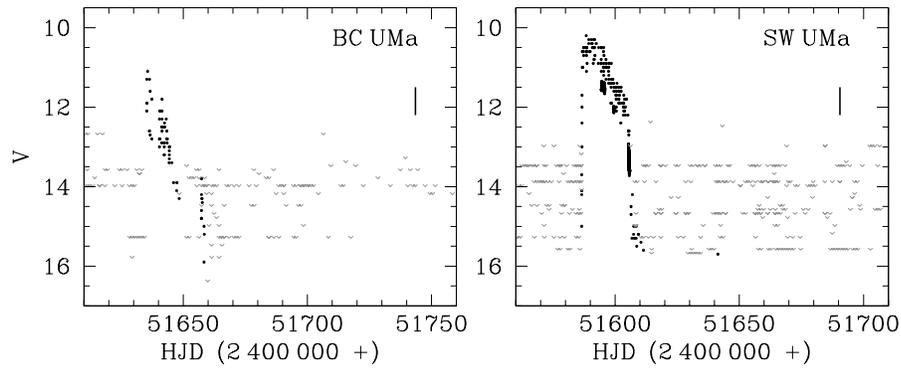}
\caption[]{\label{f-vso}
Excerpts from the AAVSO and VSNET optical long term light curves of
BC\,UMa (left) and SW\,UMa (right) close to the \textit{HST}
observations. The dates of the \textit{HST} observations are indicated
by the tick marks.}
\end{figure}

BC\,UMa was observed with STIS for five consecutive spacecraft orbits on
2000 July 18 (Table\,\ref{t-obslog}). The last outburst of BC\,UMa
previous to our STIS observations initiated on 2000 March 31, so the
system had returned to its quiescent level long before the time of our
HST pointing (Fig.\,\ref{f-vso}). Analysis of the STIS acquisition
image provides a F28$\times$50LP magnitude of $18.4$. After our
experience with the use of the E140M grating for faint objects, we
decided to use the G140L grating and the $52\arcsec\times0.2\arcsec$
aperture, providing a nominal resolution of $R\sim300$\,\kms\ over the
wavelength range 1150$-$1715\,\AA.  The STIS spectra of BC\,UMa
obtained in the five individual \textit{HST} orbits are nearly
identical and have been co-added for the subsequent analysis
(Fig.\,\ref{f-stis_opt_spectra}). As for BW\,Scl, the average STIS
spectrum of BC\,UMa also clearly reveals the photospheric spectrum of
the accreting white dwarf: a very broad \La\ profile superimposed by
the same set of narrow low-ionization metal absorption lines. The only
significant emission features are \Line{C}{III}{1176}, a broad \La\
(the narrow structures in \La\ are residuals from the not perfectly
removed geocoronal airglow) and \Line{C}{IV}{1550}.

A single optical spectrum (red/blue pair) of BC\,UMa was obtained shortly
before the \textit{HST} pointing on 3 July 2000 using the Apache Point
Observatory 3.5m telescope and Double Imaging Spectrograph with a
$1.5\,\arcsec$ slit, giving $\sim2.5$\,\AA\ resolution spectra in the
regions from $4200-5000$\,\AA\ and $6300-7300$\,\AA. The data were 
reduced and calibrated in a standard fashion using IRAF. The APO
spectrum of BC\,UMa (Fig.\,\,\ref{f-stis_opt_spectra}) is very similar to
that reported by  \citet{mukaietal90-1}, both in continuum flux and
line strength. The weak \Hb\ to \Hg\ emission lines are flanked by
broad absorption troughs, which are presumably of white dwarf
photospheric origin. 

Additional optical spectroscopy of BC\,UMa was obtained at the MMT
Observatory on the night of 29 March 2001 using the red channel
spectrograph (Table\,\ref{t-obslog}).  A total of 11 600\,s blue
(4000--5250\,\AA) spectra and a single 600\,s red (5600--8500\,\AA)
spectrum were obtained. The setup used a 1 arcsec slit giving spectral
resolutions of 1\,\AA/pixel in the blue and 0.5\,\AA/pixel in the
red. The night appeared to be nearly photometric and of constant
($\sim1.2\,\arcsec$) seeing as evidenced by watching the slit viewer
camera during the course of the observations. The data were reduced in
IRAF using average bias and red/blue flat fields taken at the start of
the night.  The standard star Feige 34, used to flux the spectra, was
observed near in time and airmass to the red spectrum and at the start
of the blue time-series sequence. The absolute flux calibration
appears good to near 10\% as estimated by using one standard star to
flux another.

Figure\,\ref{f-stis_opt_spectra} shows the average of the 11 blue
spectra along with the single red spectrum. The emission lines are
noticeably stronger and the overall flux level is higher in the MMT
data, compared to the APO spectrum and to the spectrum in
\citet{mukaietal90-1}, suggesting that the MMT observations caught the
system during a relatively active phase.  The Balmer lines are
double-peaked, suggestive of an origin from the accretion disk in a
moderately high-inclination system. As noticed by
\citet{mukaietal90-1}, the secondary star is discerned in the red part
of the spectrum. However, the quality of our single red spectrum is
not sufficient to improve the spectral typing of
\citet{mukaietal90-1}.

\subsection{SW\,UMa} 
SW\,UMa was observed with STIS during two consecutive space craft
orbits on 2000 May 26 (Table\,\ref{t-obslog}). The data were obtained
using the same instrumental setup as described in
Sect.\,\ref{s-obs_bcuma}.  The last outburst of SW\,UMa recorded
before the STIS pointing started on 2000 February 12, lasted for
$\sim20$\,d, and the system was in quiescence for $\sim85$\,d when the
\textit{HST} observations were carried out (Fig.\,\ref{f-vso}).
SW\,UMa was found at very similar flux levels during both \textit{HST}
orbits, and the average spectrum is shown in
Fig.\,\ref{f-stis_opt_spectra}. Similar to BC\,UMa, the white dwarf in
SW\,UMa is revealed through its broad \La\ absorption. However, in
SW\,UMa the emission lines are much stronger compared to BC\,UMa.  In
addition to emission of \Ion{C}{III,IV} the STIS spectrum of
SW\,UMa also contains emission of \Line{Si}{III}{1206},
\Line{N}{V}{1240}, \Line{Si}{IV}{1393,1402}, and \Line{He}{II}{1640}.
The metallic absorption lines observed in BC\,UMa are overall weaker
in SW\,UMa, possibly filled in by emission. Unambiguously detected in
absorption are the \Ion{C}{I} complex at 1280\,\AA, the
\Ion{Si}{II}/\Ion{O}{I} blend at 1300\,\AA, \Line{Ca}{II}{1432},
\Lines{Si}{II}{1527,33}, and \Line{Al}{II}{1670}.

SW\,UMa was observed on 2000 May 28 at the Apache Point Observatory.
Similar to the ultraviolet wavelength range, the optical emission
lines in SW\,UMa are also significantly stronger compared to BC\,UMa and
BW\,Scl, suggesting a stronger contribution of the accretion
disk. Correspondingly, the broad white dwarf photospheric absorption
of \Hb\ and \Hg\ is less pronounced in SW\,UMa than in BC\,UMa and
BW\,Scl.

\section{Spectral Analysis}
Our approach in analyzing and modelling the STIS data of BW\,Scl,
BC\,UMa and SW\,UMa is very similar to the methods described in detail
in our earlier papers \citep[e.g.][]{gaensickeetal01-3,
szkodyetal02-3, howelletal02-1}. Here, we will only briefly summarise our
general strategy. 

\begin{figure}
\includegraphics[angle=270,width=8.5cm]{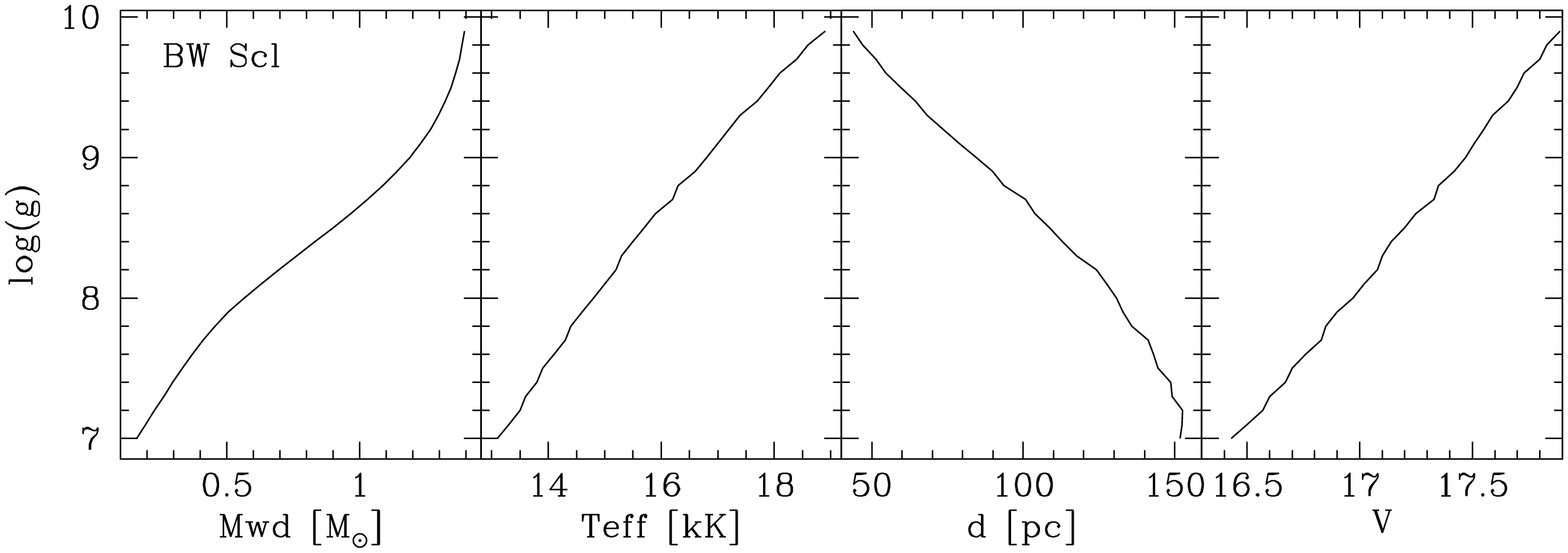}

\includegraphics[angle=270,width=8.5cm]{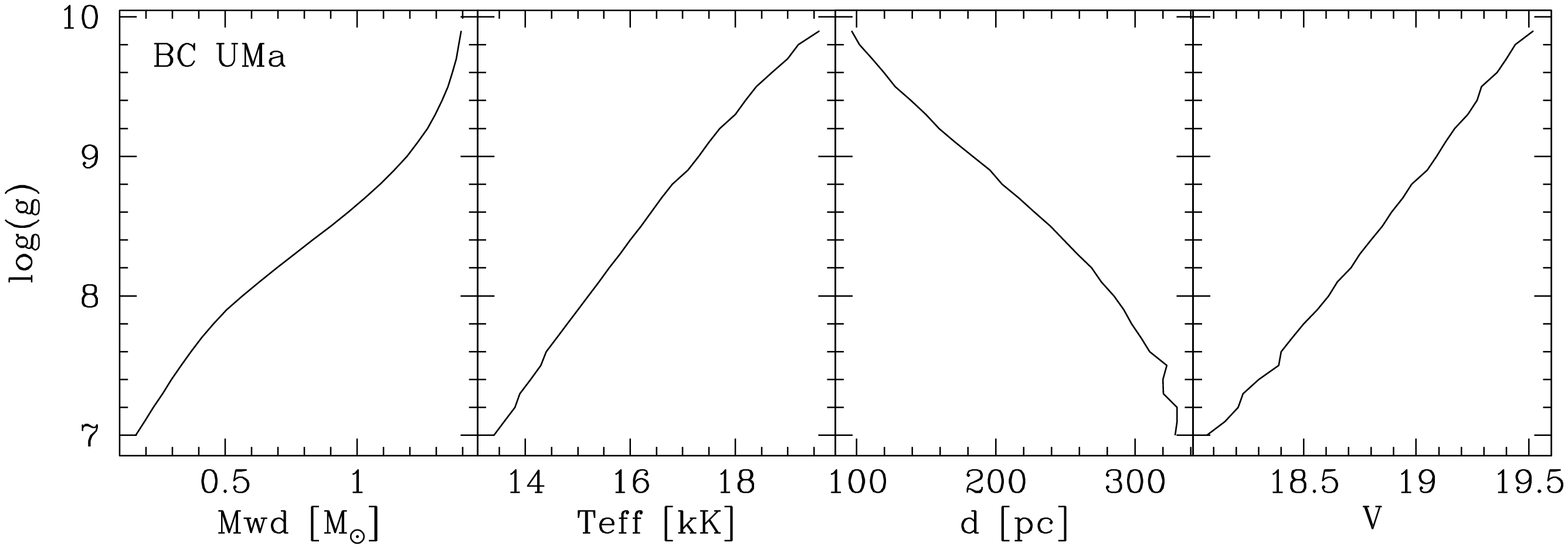}

\includegraphics[angle=270,width=8.5cm]{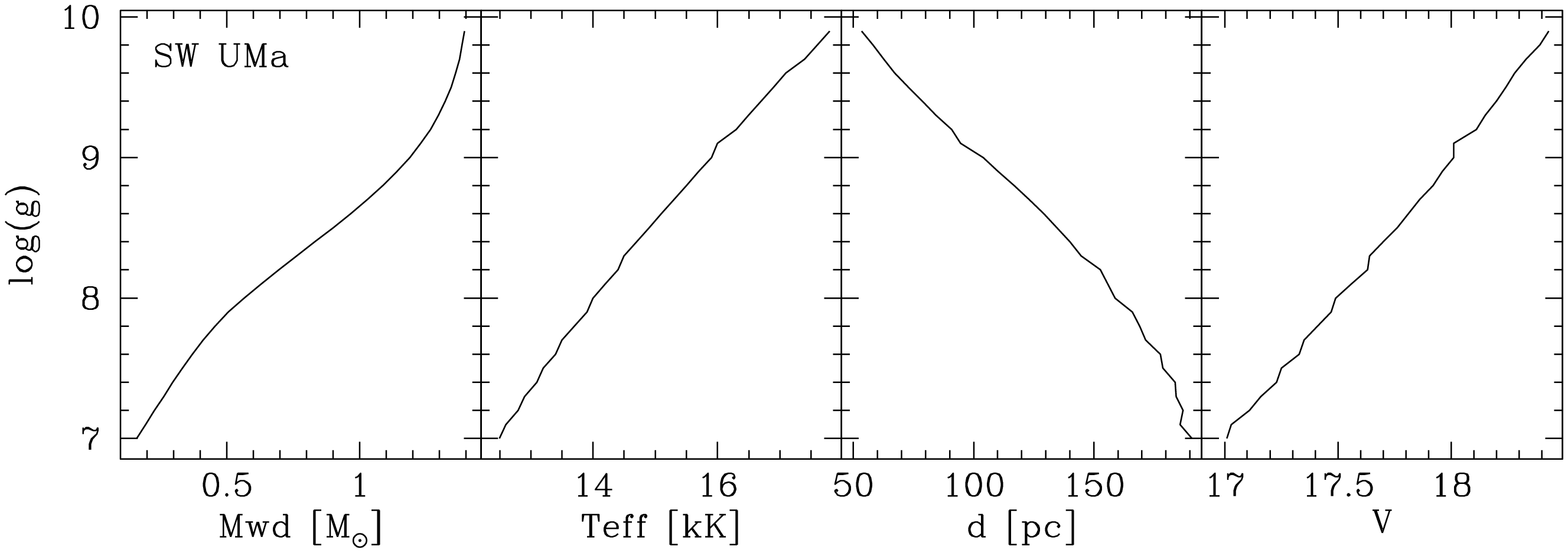}
\caption[]{\label{f-diagnostic}
Diagnostic diagrams for the white dwarf model fits to the STIS
data. The $\log g$ dependence was established by stepping through the
the grid of fixed values for $\log g$, leaving the temperature and
scaling factor free. For a given $\log g$ (or white dwarf mass), the
solid lines indicate the effective temperature, distance, and $V$
magnitude of the white dwarf implied by the fit.}
\end{figure}

At first, we derive the effective temperature \Teff\ of the white
dwarf, as well as a rough estimate of the chemical abundances in the
atmosphere. For this purpose, we use a grid of white dwarf models
calculated using TLUSTY195 and SYNSPEC45 \citep{hubeny88-1,
hubeny+lanz95-1}, which covers $\Teff=10\,000-25\,000$\,K in steps of
100\,K, $\log g=7.0-10.0$ in steps of 0.1, and metal abundances of
0.1, 0.5, and 1.0 times their solar values. Whereas fitting model
spectra to the observations of a single white dwarf allows a
determination of \Teff\ as well as the surface gravity $\log g$, the
success of this approach is very limited in the case of CV white
dwarfs for two reasons: (1) the STIS FUV data, where the white dwarf
is clearly the dominant emission component, do not provide sufficient
wavelength coverage to break the degeneracy in \Teff\ and $\log
g$. More specifically, only the red wing of \La\ is covered by the
STIS observations; (2) in contrast to single white dwarfs where
fitting the Balmer lines strongly constrains \Teff/$\log g$, the
optical spectrum of CVs is severely contaminated by emission from the
quiescent accretion disk. In the view of these limitations, we only
use \Teff\ and the metal abundances as free parameters in the fit,
keeping $\log g$ fixed. In order to account for the observed emission
lines we fit the lines with Gaussian profiles, and we exclude the
central 20\,\AA\ of \La\ from the fit to avoid contamination by
residual geocoronal \La\ emission.  The correlation between the
assumed $\log g$ (\,=\,white dwarf mass) and the best-fit value for
\Teff\ is then determined by repeating the fit while stepping through
the grid in $\log g$. A nearly linear correlation between \Teff\ and
$\log g$ is found (Fig.\,\ref{f-diagnostic}). Using a mass-radius
relation for carbon-core white dwarfs \citep{hamada+salpeter61-1} we
obtain $\Mwd(\log g)$ and $\Rwd(\log g)$. The flux scaling factor
between the white dwarf model and the STIS spectrum is then used to
calculate the distance of the system, and the $V_\mathrm{wd}$
magnitude of the white dwarf is computed. \Mwd, \Teff, $d$ and
$V_\mathrm{wd}$ are illustrated as a function of $\log g$ in the
diagnostic diagrams shown in Fig.\,\ref{f-diagnostic}.

Once that \Teff\ is established, we refine the models in terms of the
chemical abundances and white dwarf rotation rates. For each object, a
grid of model spectra is generated (adopting the best-fit temperature
for $\log g=8.0$) that covers metal abundances from 0.1 to 1.0 times
their solar values, in steps of 0.1, and rotation rates ($v \sin i$)
ranging from 200\,\kms\ to 1000\,\kms, in steps of 100\,\kms. Several
small wavelength ranges including metal lines of C, O, Si, and Al are
then fitted, normalizing the model spectra to the local continuum
(Fig.\,\ref{f-vsini}).

Additional details describing the analysis of the individual objects
are given below.

\subsection{BW\,Scl}
\begin{figure}
\includegraphics[angle=270,width=8.5cm]{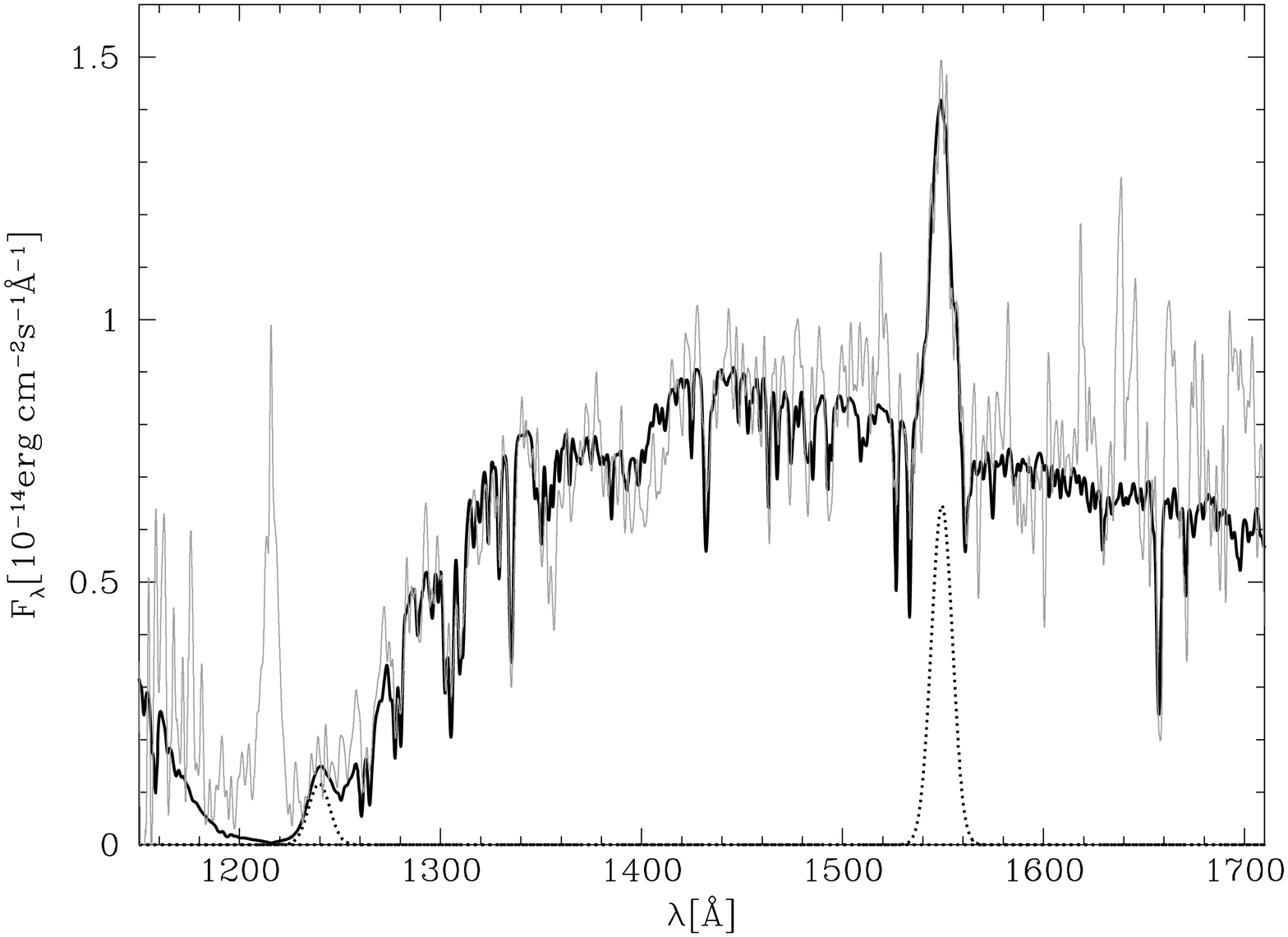}
\includegraphics[angle=270,width=8.5cm]{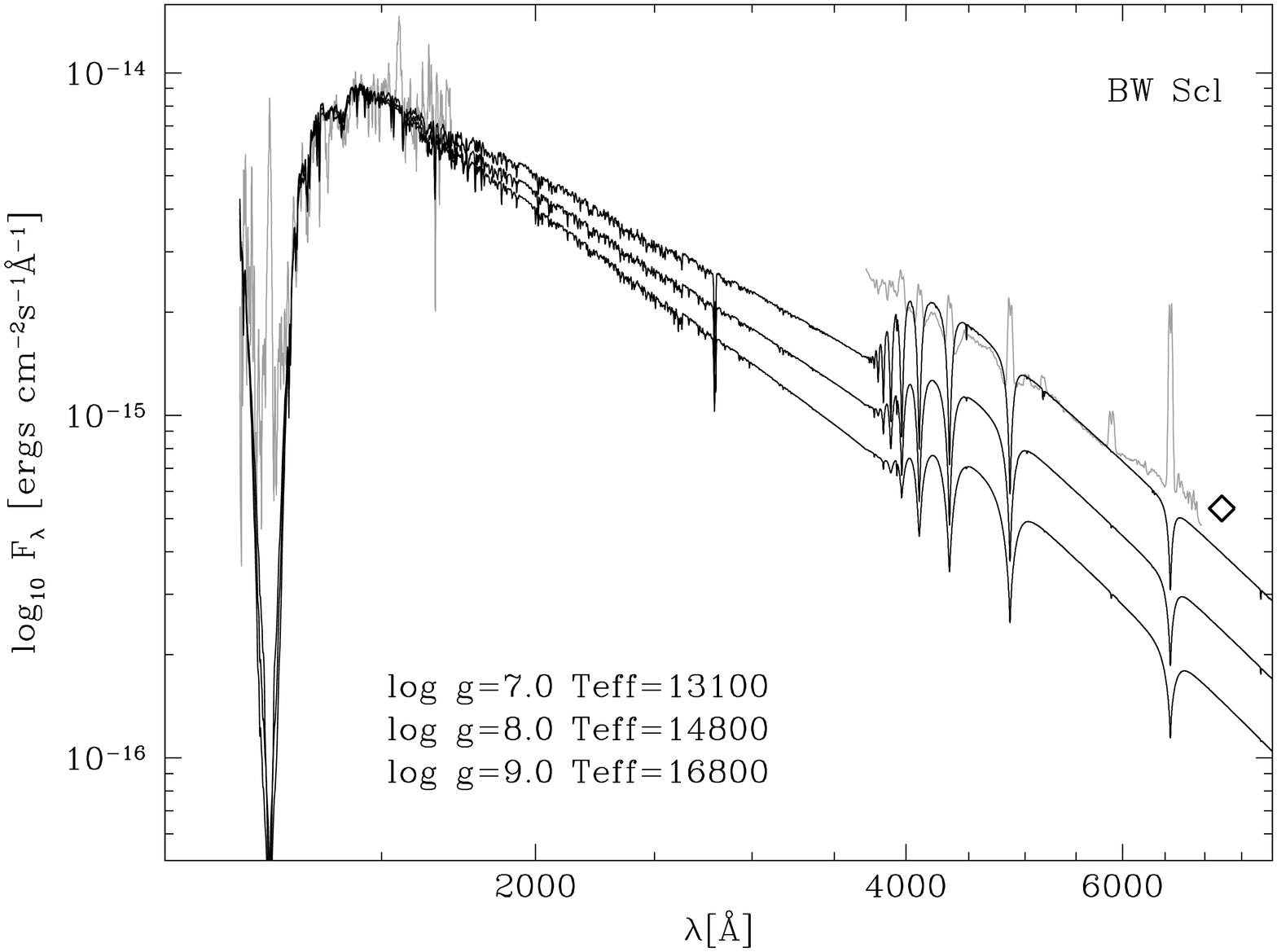}

\includegraphics[angle=270,width=8.5cm]{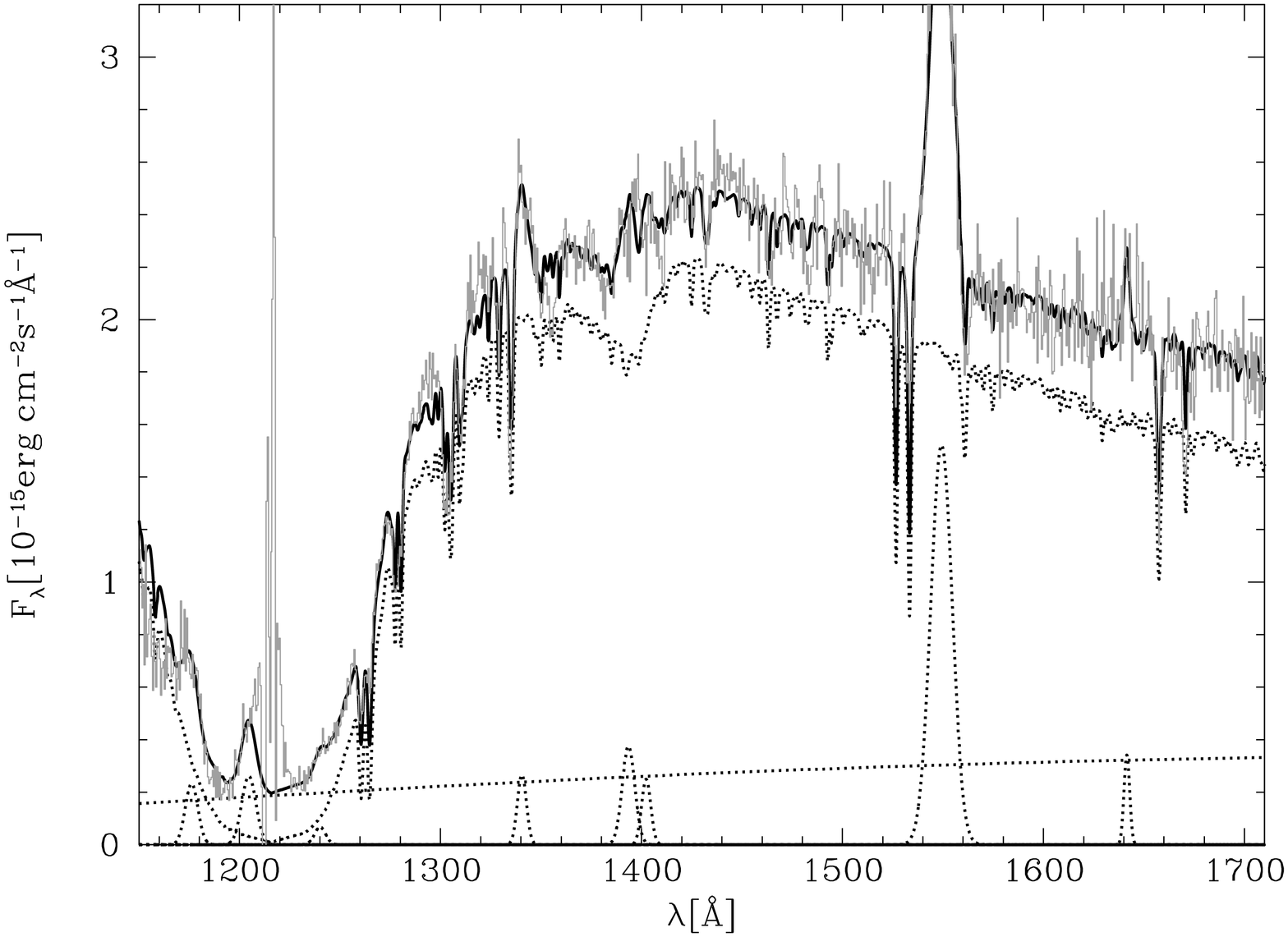}
\includegraphics[angle=270,width=8.5cm]{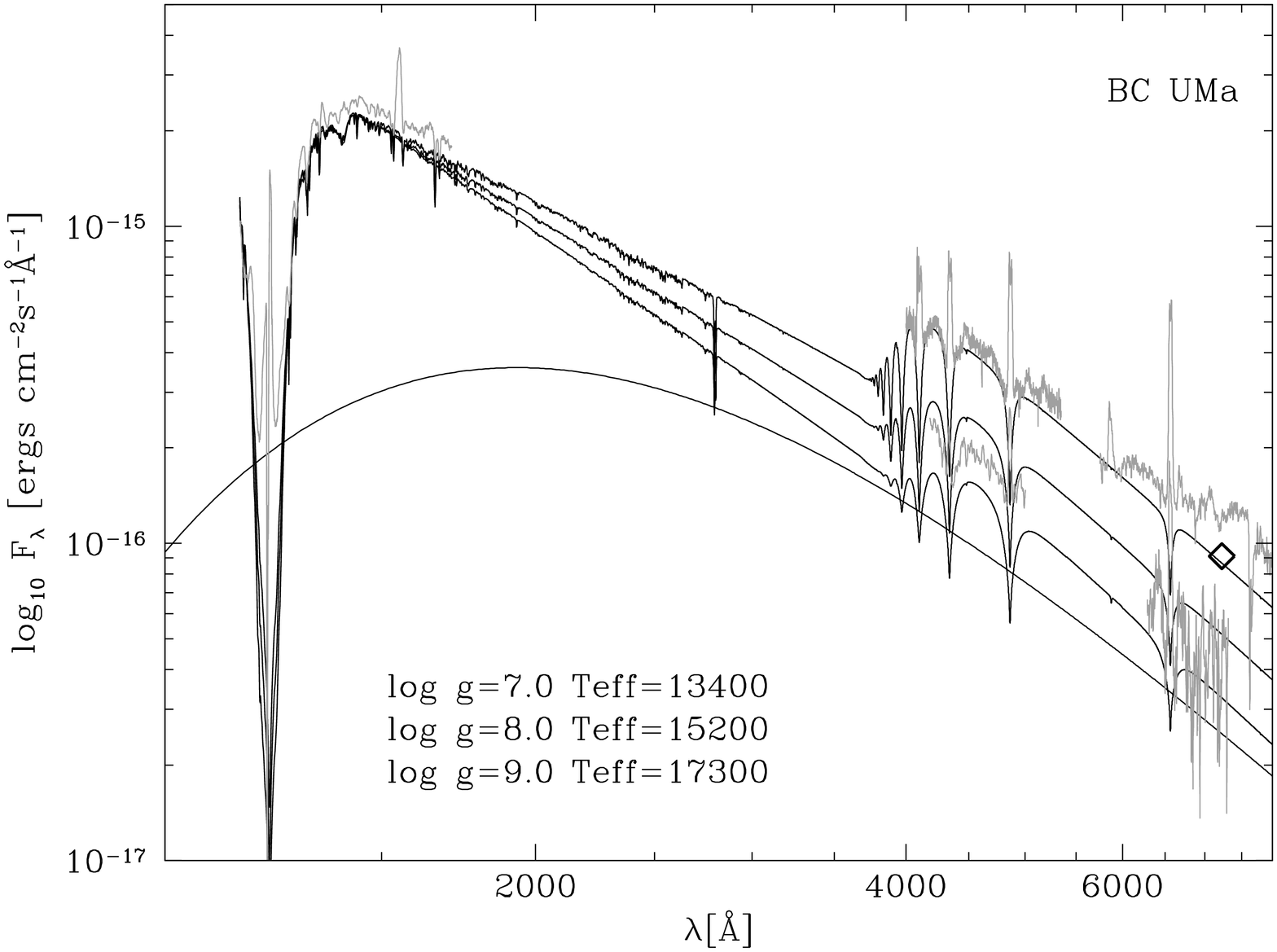}

\includegraphics[angle=270,width=8.5cm]{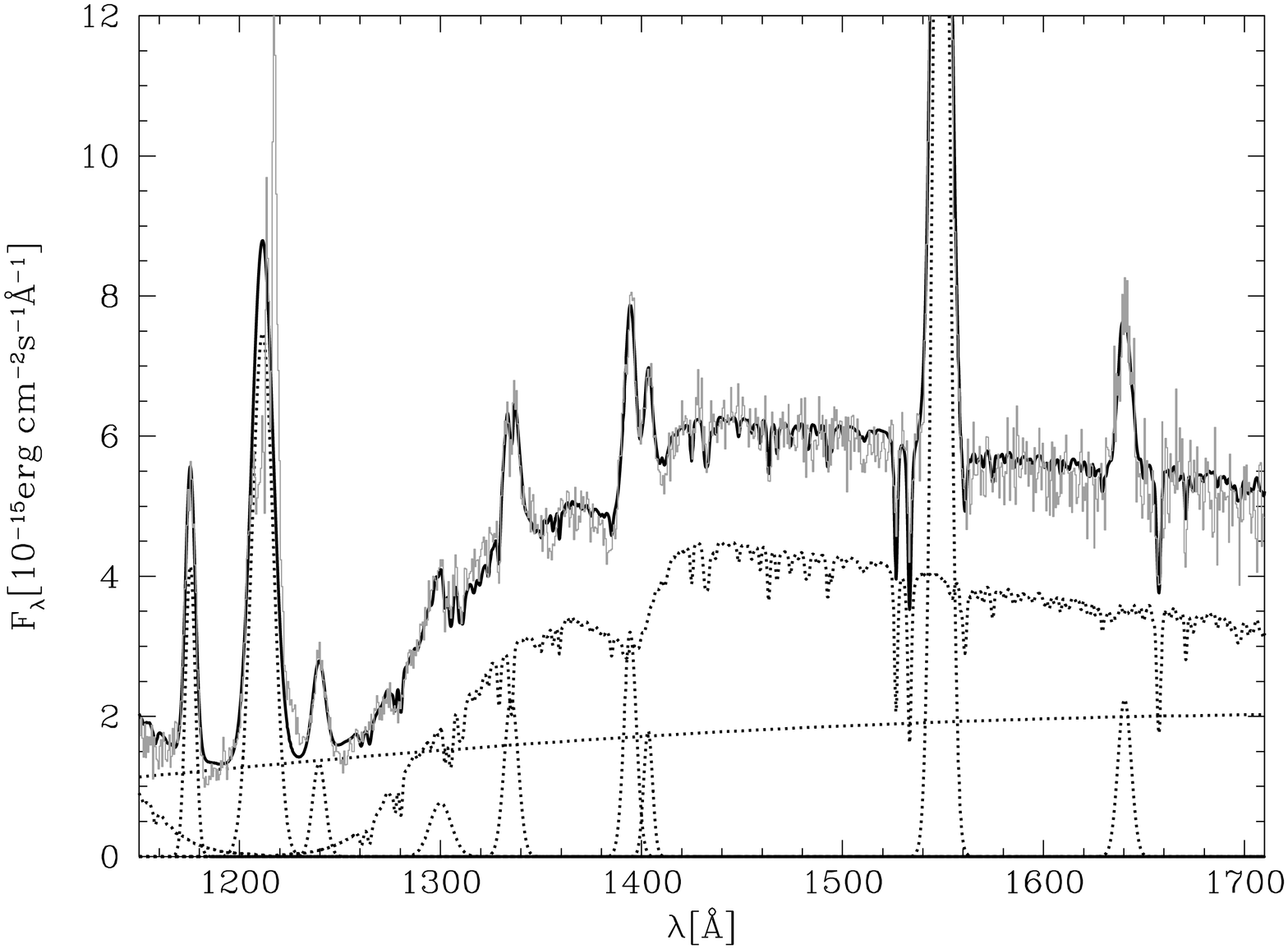}
\includegraphics[angle=270,width=8.5cm]{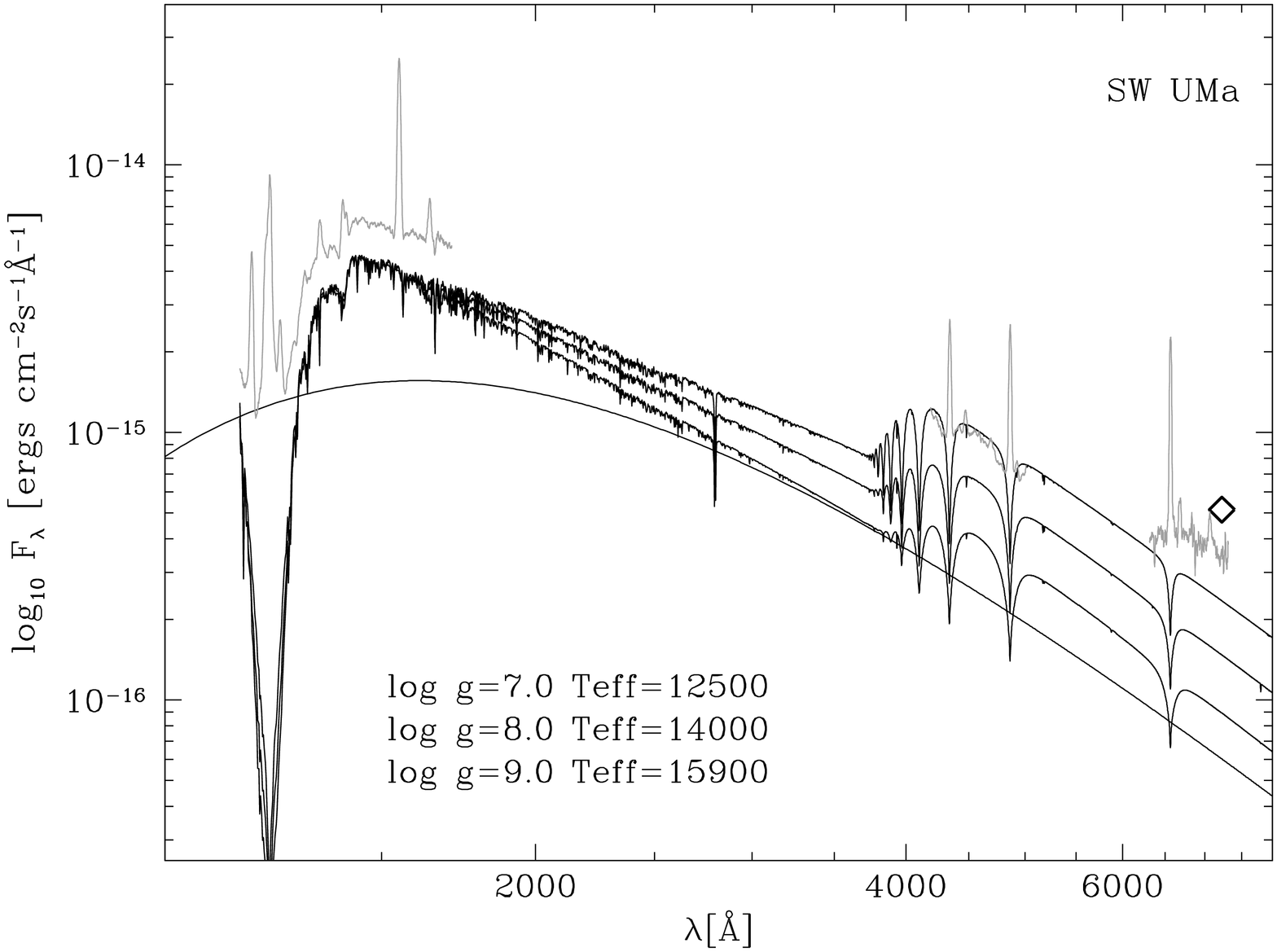}
\caption[]{\label{f-stis_fit}
Model fits to the STIS spectra. From top to bottom: BW\,Scl, BC\,UMa,
SW\,UMa. Left panels: the STIS wavelength range. For BW\,Scl, we
fitted the STIS data with a white dwarf model spectrum plus Gaussian
emission lines. BC\,UMa and SW\,UMa were modelled with a white dwarf
plus blackbody for the continuum, plus Gaussian emission lines. Right
panel: Extrapolation of the models into the optical wavelength
range. The observed spectra are shown in gray. The F28$\times$50LP
fluxes determined from the STIS acquistion images are indicated by the
open diamonds.}
\end{figure}

The STIS spectrum of BW\,Scl is well described by a
($\Teff=14\,800$\,K, $\log g=8.0$) model spectrum with 0.5 times solar
metal abundances (Fig.\,\ref{f-stis_fit}, top-left panel). We
find $\chi_\mathrm{red}^2=0.8$, which implies statistically a very good
fit. However, in view of the large errors of the echelle data
(Fig.\,\ref{f-stis_fit}), a better statement is that our model is a
good approximation considering the large uncertainties in the data.
Extending the white dwarf model spectrum into the optical range does
not provide much of a constraint on the white dwarf mass, as all
models fall below the observed spectrum, as well as below the
F28$\times$50LP magnitude derived from the STIS acquisition
image. Assuming a white dwarf mass in the range $0.35-0.90$\,\Msun\
implies $\Teff=14\,800\pm900$\,K, and a distance of $d=131\pm18$\,pc.

Fixing $\Teff=14\,800$\,K and $\log g=8.0$, we determine $\simeq0.5$
times solar abundances for carbon, oxygen and silicon, whereas
aluminium is found to be significantly overabundant (by a factor
three) with respect to its solar value
(Fig.\,\ref{f-vsini}). We have investigated whether the effect
of interstellar absorption could mimic the derived Al
overabundance. Inspection of FUV interstellar absorption lines at both
low and high galactic latitudes shows that the strength of
\Line{Al}{II}{1670} is comparable to that of \Line{Si}{II}{1260, 1302,
1527}, \Line{O}{I}{1302}, and \Line{C}{II}{1335}
\citep[e.g.][]{lehneretal01-1, welshetal01-1}. As we do not detect
excess absorption in any of these other lines, we conclude that the
observed enhancement of the Aluminium line is intrinsic to BW\,Scl,
and suggests a substantial overabundance of Al in the photosphere of
the white dwarf. The only caveat to this statement is that
\Line{Al}{II}{1670} is by far the strongest (factor $\sim10$) Al line
in the entire ultraviolet wavelength range. Confirming the Al
abundance derived here would need high-resolution FUV (e.g. E140M)
data at high S/N to make use of $\sim10$ weaker \Ion{Al}{II,III} lines
for spectral modelling. The narrow widths of the observed metal
absorption lines implies a low rotation rate, $v\sin i<300$\,\kms.

\subsection{\label{s-bcuma} BC\,UMa}
Applying a simple white dwarf model fit to the STIS observations of
BC\,UMa results in $\Teff=15\,400$\,K for $\log g=8.0$, however, this
fit does not adequately reproduce the wavelength range
$\simeq1180-1250$\,\AA, where the observed flux level significantly
exceeds that of the model spectrum. In order to account for this
additional component, we have repeated the fits to the STIS data of
BC\,UMa, allowing for either a power-law or a blackbody contribution
to the FUV spectrum. The best-fit parameter for the second component
is either a power-law index of $\alpha=-0.04$ or a blackbody
temperature of 14\,000\,K. In both cases, the additional continuum
component contributes $\simeq10\%$ of the flux at 1400\,\AA. The white
dwarf effective temperature remains nearly unchanged,
$\Teff=15\,200$\,K, but the somewhat smaller scaling factor results in
a slightly larger distance estimate. Assuming again a white dwarf mass
in the range $0.35-0.90$\,\Msun\ we find $\Teff=15\,200\pm1000$\,K,
and a distance of $d=285\pm42$\,pc.

Adding the second continuum component improves the fit in
terms of $\chi_\mathrm{red}^2$, which drops from $\simeq3.4$ for the
white-dwarf-only fit to $\simeq2.2$ for the two-component models. On
statistical grounds, both fits are poor, which is a sign that either
the flux errors provided by the STIS pipeline are underestimated;
there are additional systematic errors not accounted for by the
pipeline; or that our model is not fully appropriate to describe the
data. The reality is likely to be a mix of the second and third
possibility. We have inspected high S/N  STIS G140L
spectra of the $\simeq20\,000$\,K single DA white dwarf
GRW+70$^\circ$5824 which reveal various wiggles that are not real and
result in $\chi_\mathrm{red}^2>1$ when fit with a model spectrum. In
addition, our simple approach for approximating the disk emission by a
component made up from a smooth slope plus Gaussian emission lines is
certainly a fairly poor model for the reality. Keeping these
limitations in mind, we adopted the following strategy to assess the
statistical relevance of the second continuum component. We have
renormalized the errors of the observed STIS spectrum so that
$\chi_\mathrm{red}^2=1$ for the two-component model. The wavelength
independent factor used for this procedure was 1.45, which, assuming
we had a perfect model for the data and no additional systematic
errors, would imply that the flux errors provided by the pipeline are
underestimated by this factor. Next, we calculated
$\chi_\mathrm{red}^2\simeq1420/903\simeq1.57$ for the single white
dwarf fit using the renormalized errors, which implies that the single
white dwarf fit is an inacceptable description of the data.
On statistical grounds there is, however, no preference for either the
white dwarf plus blackbody or white dwarf plus power law model: both
two-component fits provide basically a nearly flat continuum component
underlying the white dwarf emission, and improve the fit by similar
amounts with respect to a single white dwarf model. The physical
interpretation of the additional continuum component is emission
either associated with the quiescent accretion disk or the hot
spot. In the case of optically thin emission from an optically thin
disk (or a thin hot layer on top of a colder optically thick disk), a
power-law description would account for the wavelength dependence of
the free-free absorption coefficient. Our blackbody approach would be
a simple description for optically thick emission from the hot
spot. At a distance of $d=285$\,pc, the radius of the blackbody
component, assuming a spherical shape, would be
$\simeq2.9\times10^8$\,cm. The temperature implied by the fit appears
somewhat high with respect to the existing temperature estimates for
hot spots (e.g. \citealt{panek+holm84-1, mateo+szkody84-1,
szkody87-2}), which are mostly based on blackbody analyses of either
the orbital flux variation in the ultraviolet and/or optical or of the
eclipse ingress/egress profiles. It is not too surprising that fitting
a blackbody to the FUV continuum results in a different temperature
estimate, moreover a blackbody is only a very simple
approach. Extending the second component into the optical, it is clear
that a simple power-law is an unphysical concept, as the free-free
absorption coefficient breaks down at the Balmer edge. In
Fig.\,\ref{f-stis_fit} we have plotted the UV-optical spectral energy
distribution of BC\,UMa, along with the best-fit white dwarf plus
blackbody model. Taking the flux level of the APO spectra at face
value, they would constrain $\log g>8.0$,
i.e. $\Mwd>0.6$\,\Msun. However, the F28$\times$50LP flux at the time
of the \textit{HST} observations exceeds that of the APO spectra, and
we remain without a stringent constraint on the white dwarf mass.
Real progress on the SED modelling of quiescent dwarf novae
will need a realistic physical model for the emission of the accretion
disk.

For the detailed metal line fitting, we fixed $\Teff=15\,200$\,K and
$\log g=8.0$, and find carbon, oxygen and silicon, abundances at
$0.3\pm0.1$ times their solar values. As for BW\,Scl, the fit to
\Line{Al}{II}{1670} requires the abundance of aluminium to be
significantly (by a factor two) super-solar. In contrast to
the E140M echelle data of BW\,Scl, the G140L of BC\,UMa is not
affected by instrumental features in the region around
\Line{Al}{II}{1670}, lending further support for the detection of an
enhanced Aluminium abundance.
The observed line profiles
are best fit with a rotation rate of $v\sin i=300\pm100$\,\kms. 

\subsection{SW\,UMa}
In the case of SW\,UMa, the additional contribution in excess of the
white dwarf photosphere is even more evident than in BC\,UMa. The
emission lines of \Line{C}{III}{1176}, \Lines{C}{IV}{1550} and
\Line{He}{II}{1640} are much stronger,  the emission lines of
\Line{Si}{III}{1206}, \Line{N}{V}{1240}, \Line{C}{II}{1335},
\Lines{Si}{IV}{1393,1402} are also clearly present, and the continuum
flux underlying the white dwarf spectrum is substantially stronger. 

Fitting the STIS spectrum with a white dwarf alone (plus Gaussian
profiles for all the detected emission lines) results in
$\Teff=14\,700$\,K for $\log g=8.0$, but with a very high
$\chi_\mathrm{red}^2$ of $\sim9.6$. Consequently, we applied the
two-component model described in Sect.\,\ref{s-bcuma} to the STIS spectrum
of SW\,UMa and find $\Teff=13\,900$\,K for the white dwarf (again for
$\log g=8.0$) plus a blackbody of $T=17\,000$\,K or a power law with a
spectral index $\alpha=-0.04$ (Fig.\,\ref{f-stis_fit}). The non-white
dwarf contribution to the continuum flux is $\simeq20$\,\% at
1400\,\AA. For a white dwarf mass in the range $0.35-0.90$\,\Msun, our
fit to the STIS data implies $\Teff=13\,900\pm900$\,K and
$d=159\pm22$\,pc. Adopting this two-component approach makes a
substantial difference in $\chi^2_\mathrm{red}$, which drops to
$\simeq2.4$ for both cases, white dwarf plus either a blackbody or
a power law.  The temperature of the blackbody component seems too
high for a hot spot, but as mentioned above, a blackbody is probably
an oversimplified model for the hot spot~--~if the hot spot is indeed
the source of the observed additional continuum flux. For completeness
the radius of the blackbody component, assuming spherical geometry and
a distance of $d=159$\,pc, is $3.4\times10^8$\,cm~--~very similar to
that found in the case of BC\,UMa. For SW\,UMa, the optical spectra do
not provide an additional constraint on $\log g$, as all white models
fall below the optical flux.

Fixing $\Teff=13\,900$\,K and $\log g=8.0$, we refined the analysis of
the metal absorption lines, and find the abundances of carbon, oxygen
and silicon to be at $0.3\pm0.1$ times their solar values.  Similar to
BW\,Scl and BC\,UMa, the abundance of aluminium is supra-solar by a
factor 1.7.  Some care has to be applied to these abundance
measurements, as the presence of various strong emission lines could
imply that the absorption lines observed in the STIS spectrum are
filled in to some extent by emission. This should, however, not affect
the large abundance ratio of aluminium to the other elements. It
appears that the metal lines are not resolved at the resolution of the
G140L grating, implying $v\sin i<300$\,\kms.

\section{Short-term variability}
All STIS data were obtained in the TIME-TAG mode, recording the
arrival time and detector coordinate for each individual photon. This
enables us to extract light curves over any selected wavelength range
and at an arbitrary time resolution. 

In the case of BW\,Scl, however, the use of the E140M echelle grating, which
distributes the source photons over 44 individual echelle orders on
the MAMA detector, prevents a reliable background
subtraction. Extracting all photons and binning them into a 30sec
light curve shows a low-amplitude modulation on a time scale of
$\sim30$\,min. Alas, with the unknown contribution of the
background, it is not possible to unequivocally ascribe this
modulation to the source, or to the variation of the background
throughout the orbit of \textit{HST}. However, the STIS TIME-TAG data
do not show any evidence for short-period flickering. 

BC\,UMa was observed with the G140L first-order grating, and
background-subtracted light curves were extracted for both the
continuum and the \Ion{C}{IV} line (see \citealt{gaensickeetal01-1}
for a description of the methods used). Within the errors, the count
rates are consistent with constant emission.

\begin{figure*}
\includegraphics[angle=270,width=14cm]{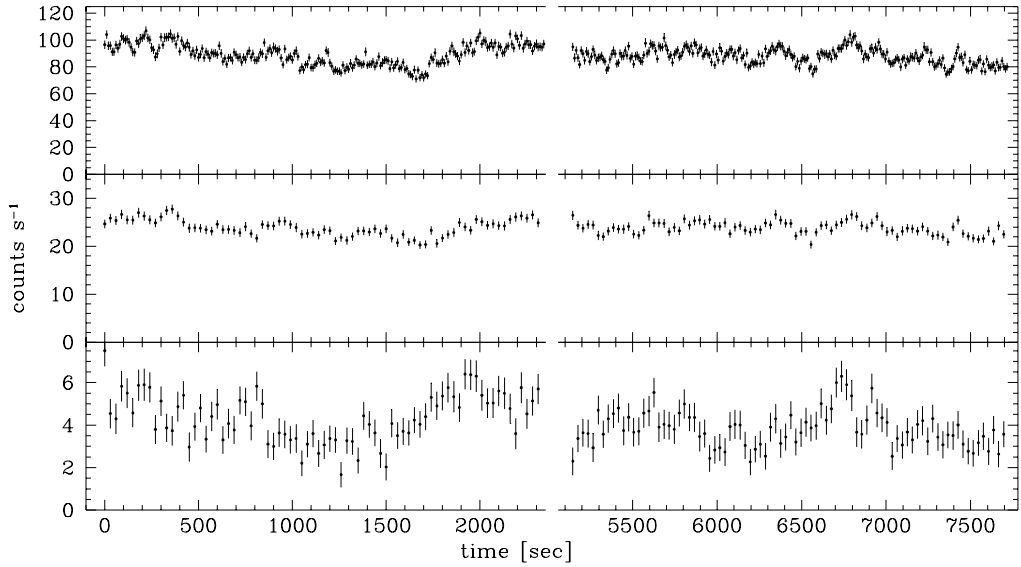}
\caption[]{\label{f-ttag}
The far-ultraviolet light curves of SW\,UMa extracted from the STIS
TIME-TAG data. Top panel: $1230-1710$\,\AA, 10\,s time
resolution. Middle panel: line-free continuum, $1420-1520$\,\AA, 30\,s
time resolution. Bottom panel: \Ion{C}{IV} line, 30\,s time resolution.}
\end{figure*}

SW\,UMa was also been observed with the G140L grating, and we extracted
continuum and \Ion{C}{IV} light curves
(Fig.\,\ref{f-ttag}). Variability is seen on time scales ranging from
$\sim100$\,sec to $\sim1500$\,sec. A period (Lomb-Scargle) analysis of
the data does not reveal any periodicities, specifically no signal is
found either at the hypothetical 15.9\,min white dwarf spin period
\citep{shafteretal86-2}, or at the orbital period. The variance of the
continuum and \Ion{C}{IV} light curves are 6\% and 25\%, respectively,
clearly indicating that the flickering is primarily associated with
the emission lines~--~in agreement with our results on EK\,TrA
\citep{gaensickeetal01-3}.

\section{Discussion \& Conclusions}
As part of our medium-size Cycle\,8 \textit{HST}/STIS program, we have
observed 16 cataclysmic variables, with the aim of significantly
improving our knowledge about the properties of accreting white dwarfs
in cataclysmic variables \citep{szkodyetal02-1}. For twelve systems,
our strategy/selection worked well, and the analysis of the STIS
spectra of their white dwarfs roughly doubled the number of effective
temperature measurements for non-magnetic CV white dwarfs, and
quadrupled the number of abundance/rotation rate
estimates\footnote{The dwarf novae TY\,Psc, TZ\,Per were observed in
outburst, DI\,Lac is an old nova where the white dwarf could not be
detected \citep{moyeretal03-1}, and the analysis of the helium CV
CP\,Eri is still in progress.}.

The white dwarf temperatures found range from $\simeq12\,300$\,K
(EG\,Cnc, \citealt{szkodyetal02-3}) to $\simeq18\,800$\,K (EK\,TrA,
\citealt{gaensickeetal01-3}), with a strong clustering near
$\simeq15\,000$\,K~--~where the three systems presented here also
fall. In a simple-minded picture of CV evolution, one might expect the
lowest white dwarf temperatures near the minimum period, where the
accretion rates should be lowest. In our sample, this is not the case:
BW\,Scl is the shortest-period ($\Porb=78.2$\,min) dwarf nova with
$\Teff\simeq14\,800$\,K, whereas the coldest white dwarf
($\Teff\simeq12\,300$\,K) is found in EG\,Cnc at
$\Porb=86.4$\,min. However, one significantly colder white dwarf has
recently been found near the minimum period, HS2331+3905
($\Porb=81.1$\,min, $\Teff\simeq10\,500$\,K;
\citealt{araujo-betancoretal05-1}), and some of the CV white dwarfs
discovered by Sloan \citep{szkodyetal02-2,szkodyetal03-2} may turn out
to be very cold as well. The observed spread in $\Teff$ might to some
extent be related to different white dwarf masses, as the accretion
luminosity is, at constant accretion rate, proportional to
$\Mwd/\Rwd$.

None of the 12 white dwarfs observed within our program is a rapid
rotator~--~in fact, for most systems we could only determine upper
limits on the rotation rate. This strongly suggests that the angular
momentum coupling between the white dwarf envelope and its core is
rather inefficient, and that the bulk of the accreted angular momentum
is lost in the shell ejected during nova eruptions.

We have found substantial sub-solar abundances in all 12 white
dwarfs analyzed in this program. It appears unlikely that these low
white dwarf photospheric abundances reflect significant sub-solar
metallicities of their donor stars, as the majority of CVs should be
ordinary galactic disk objects. It is more likely that the abundances
observed in the white dwarf photospheres reflect an
accretion/diffusion equilibrium (radiative levitation is irrelevant
for the low temperatures of most CV white dwarfs). While it would be
of fundamental importance to estimate accretion rates from such
equilibrium abundances, a quantitative exploitation of this method is
currently not possible as all published diffusion calculations have
been carried out either for single DA or single DB white dwarfs, and
we underline the need for detailed studies of diffusion time scales
for various ions in CV white dwarf atmospheres. An additional
uncertainty in the interpretation of the sub-solar abundances in CV
white dwarfs is the uncertain distribution of the accreted material
over the white dwarf surface, i.e. whether a lateral abundance
gradient from the equator, where the material is accreted from the
accretion disk, to the poles exist. While the current data can not be
used to assess this possibility, future spectroscopy with high
resolution and high S/N might shed light on this question, as the
detailed shape of Doppler-broadened line profiles will depend on such
a lateral abundance gradient.

Unfortunately, the low S/N prevented detailed abundance
measurements of individual chemical species for most systems.  The
high S/N data presented here suggest an overabundance of aluminium in
BW\,Scl, BC\,UMa, and SW\,UMa. At present, we are not yet able to
interpret the abundances anomalies of some elements that have been
observed in a number of CV white dwarfs in an unambiguous way.  The
peculiar abundances of odd-numbered nuclei may be related to
proton-capture during classical nova eruptions, as suggested by
\citet{sionetal97-1}. Alternatively, abundance anomalies seen in the
white dwarf atmosphere may just reflect abundance anomalies of the
companion star as a result of its evolution in a binary system
\citep{gaensickeetal03-1, harrisonetal04-1}.

Our Cycle\,8 program has quadrupled the number of reliable effective
temperature measurements for short-period dwarf novae, and
demonstrated the potential for abundance and rotation velocity
measurements.  Further progress in our understanding of both the
properties of white dwarfs in CVs as well as CV evolution needs a
larger sample of well-observed systems \textit{and} a reliable
estimate of their white dwarf masses~--~however, this progress
crucially depends on the availability of FUV instruments with a higher
throughput and higher spectral resolution, such as the Cosmic Origin
Spectrograph could provide. 

\section*{Acknowledgements}
Based on observations made with the NASA/ESA Hubble Space Telescope,
obtained at the Space Telescope Science Institute, which is operated
by the Association of Universities for Research in Astronomy, Inc.,
under NASA contract NAS 5-26555.  Part of the optical observations
reported here were obtained at the MMT Observatory, a joint facility
of the University of Arizona and the Smithsonian Institution.  BTG was
supported by a PPARC Advanced Fellowship. PS, EMS, and SBH acknowledge
partial support of this research from HST grant GO-08103.03-97A. SBH
acknowledges partial support of this work from NSF grant AST
98-10770. An anonymous referee contributed a number of useful comments
which helped to improve the paper.


\newpage

\begin{deluxetable}{lccccc}
\tablecolumns{5}  
\tablewidth{0pc}  
\tablecaption{\label{t-obslog}Log of the observations}
\tablehead{  
\colhead{Object} &
\colhead{Telescope} &
\colhead{Grating/} &
\colhead{Date}  &
\colhead{Magnitude}  &
\colhead{Exposure time} \\
\colhead{} &
\colhead{} &
\colhead{Resolution} &
\colhead{UT} &
\colhead{} &
\colhead{[s]}}
\startdata  
BW\,Scl & HST & E140M    & 1999-09-12 & 16.5$^a$ & 1977s \\
        & ESO\,3.6m$^\dag$ & 16\,\AA & 1992-10-01 & 16.5$^b$ & 600\,s \\
BC\,UMa & HST & G140L    & 2000-07-18 & 18.4$^a$ & 12998s \\
        & APO & 2.5\,\AA & 2000-07-03 & 18.7$^b$ & 600s   \\
        & MMT & 2.0\,\AA & 2001-03-29 & 18.0$^b$ & $11\times600$\,s\\
SW\,UMa & HST & G140L    & 2000-05-26 & 16.5$^a$ & 4933s \\
        & APO & 2.5\,\AA & 2000-05-28 & 16.9$^b$  & 600s   \\
\multicolumn{6}{l}{$^a$ F28x50LP ~~~~~ $^b$ $V$ ~~~~~ $^\dag$ from \citep{abbottetal97-1}}
\enddata 
\end{deluxetable}

\begin{deluxetable}{lcccc}
\tablecolumns{5}  
\tablewidth{0pc}  
\tablecaption{\label{t-fitpar}Best-fit white dwarf parameter assuming
  $\log g=8.0\pm0.5$}
\tablehead{  
\colhead{Object} &
\colhead{\Teff\,[K]} &
\colhead{$v\sin i$\,[\kms]}  &
\colhead{Metal Abundances $(\odot)$} &
\colhead{$d$\,[pc]} }
\startdata  
BW\,Scl & $14\,800\pm900$ & $<300$      & Al\,$3.0\pm0.8$, rest $0.5\pm0.2$ & $131\pm18$\\
BC\,UMa & $15\,200\pm1000$ & $300\pm100$ & Al\,$2.0\pm0.5$, rest $0.3\pm0.1$ & $285\pm42$\\
SW\,UMa & $13\,900\pm900$ & $<300$      & Al\,$1.7\pm0.5$, rest $0.2\pm0.1$ & $159\pm22$\\
\enddata  
\end{deluxetable}

\clearpage

\clearpage

\clearpage

\begin{figure}
\includegraphics[angle=270,width=12cm]{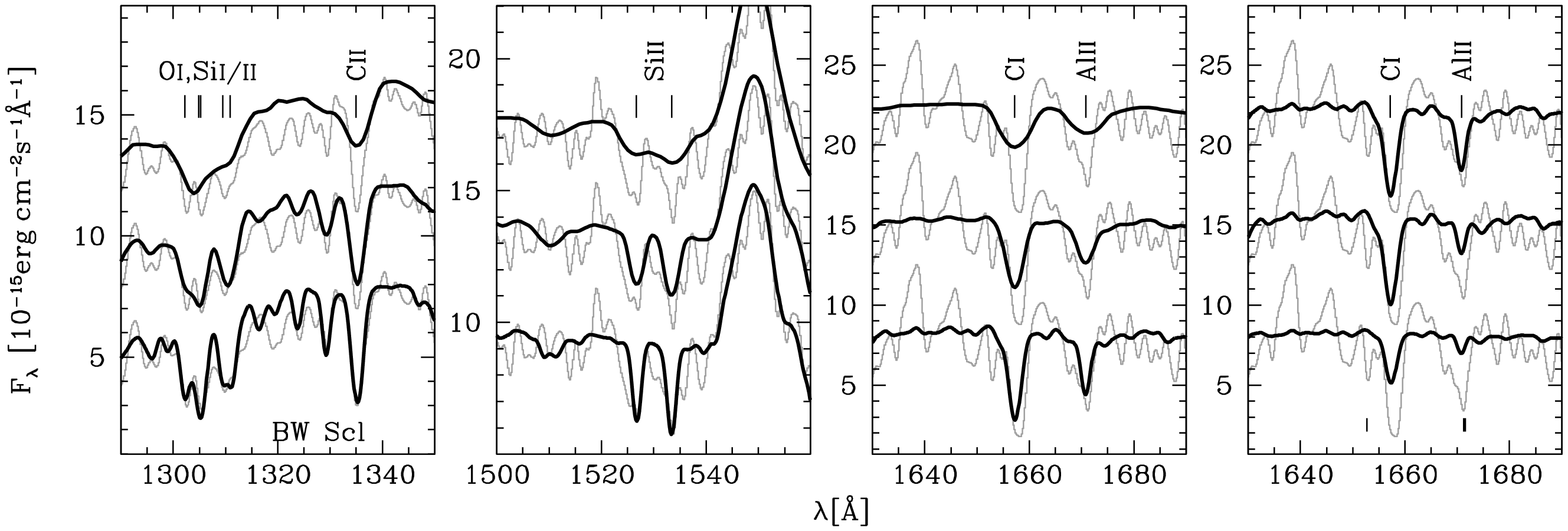}

\includegraphics[angle=270,width=12cm]{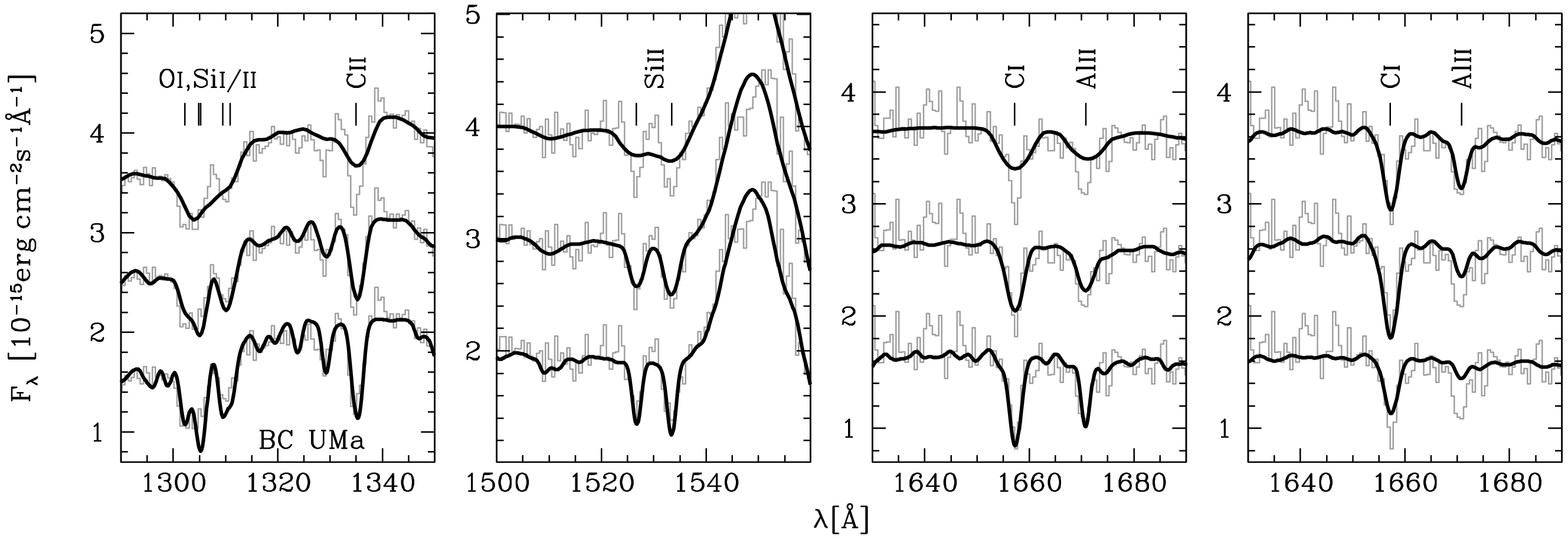}

\includegraphics[angle=270,width=12cm]{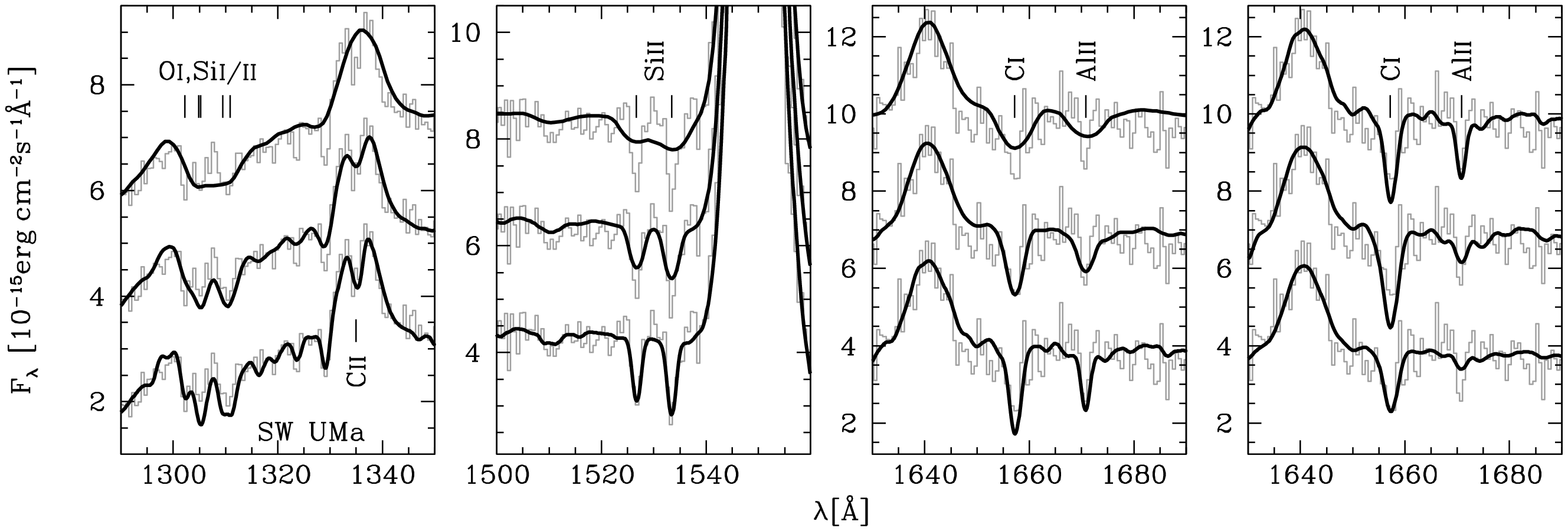}
\caption[]{\label{f-vsini}
Detailed fits to the absorption lines of carbon, oxygen, silicon and
aluminium. Top panels: BW\,Scl, all models have $\Teff=14\,800$\,K and
$\log g$. The three left panels show the STIS data along with with
white dwarf model spectra for $0.5\times$ solar abundances and
rotation rates of 200\,\kms\ (bottom curve), 400\,\kms\ (middle curve)
and 800\,\kms\ (top curve). The right panel shows models for a
rotation rate of 200\,\kms\ and $0.1\times$ solar abundances (bottom),
$0.7\times$ solar abundances (middle) and $0.3\times$ solar abundances
for all metals except $3.0\times$ solar abundances for
aluminium. Middle panels: equivalent plots for BC\,UMa, with
$\Teff=15\,100$\,K and $\log g=8.0$. The abundances are $0.3\times$
solar in the left panels where varies from $v\sin i=200, 400,
800$\,\kms. In the right panel, $v\sin i=300$\,\kms, but $0.1\times$
solar abundances in the bottom curve, $0.6\times$ solar in the middle
curve. The top curve has $0.3\times$ solar for all metals except
aluminium, which has $2.0\times$ solar abundances. Bottom panels:
equivalent plots for SW\,UMa, with $\Teff=13\,900$\,K and $\log
g=8.0$. The metal abundances in the left panels are $0.2\times$ solar
their values, and $v\sin i=200, 400, 800$\,\kms. The right panel shows
$v\sin i=300$\,\kms, and $0.1\times$ solar abundances (bottom curve)
and $0.5\times$ solar abundances (middle curve). The top curve shows
all metals at $0.2\times$ solar abundances, except aluminium at
$1.7\times$ its solar value}
\end{figure}

\clearpage

\clearpage

\end{document}